\documentclass[preprint]{aastex61}
\usepackage{epstopdf}
\usepackage{color}
\usepackage{amssymb}

\usepackage{pbox}
\usepackage{graphicx}

\received{\today}
\submitjournal{ApJ}

\shorttitle{Flares\&Substorms}
\shortauthors{Artemyev et al.}

\begin{document}

\title{Comparative study of electric currents and energetic particle fluxes in a solar flare and Earth magnetospheric substorm}  

\correspondingauthor{Anton V. Artemyev}
\email{aartemyev@igpp.ucla.edu}

\author{Anton Artemyev}
\affiliation{Department of Earth, Planetary, and Space Sciences and Institute of Geophysics and Planetary Physics, \\University of California, Los Angeles, California, USA}
\affiliation{Space Research Institute, RAS, Moscow, Russia}

\author{Ivan Zimovets}
\affiliation{Space Research Institute, RAS, Moscow, Russia}

\author{Ivan Sharykin}
\affiliation{Space Research Institute, RAS, Moscow, Russia}

\author{Yukitoshi Nishimura}
\affiliation{Department of Electrical and Computer Engineering and Center for Space Sciences, Boston University, Boston, MA USA}

\author{Cooper Downs}
\affiliation{Predictive Science Inc., San Diego, CA, United States}

\author{James Weygand}
\affiliation{Department of Earth, Planetary, and Space Sciences and Institute of Geophysics and Planetary Physics, \\University of California, Los Angeles, California, USA}

\author{Robyn Fiori}
\affiliation{Geomagnetic Laboratory, Natural Resources Canada, Ontario, Ottawa, Canada}

\author{Xiao-Jia Zhang}
\affiliation{Department of Earth, Planetary, and Space Sciences and Institute of Geophysics and Planetary Physics, \\University of California, Los Angeles, California, USA}

\author{Andrei Runov}
\affiliation{Department of Earth, Planetary, and Space Sciences and Institute of Geophysics and Planetary Physics, \\University of California, Los Angeles, California, USA}

\author{Marco Velli}
\affiliation{Department of Earth, Planetary, and Space Sciences and Institute of Geophysics and Planetary Physics, \\University of California, Los Angeles, California, USA}

\author{Vassilis Angelopoulos}
\affiliation{Department of Earth, Planetary, and Space Sciences and Institute of Geophysics and Planetary Physics, \\University of California, Los Angeles, California, USA}

\author{Olga Panasenco}
\affiliation{Advanced Heliophysics, Pasadena, CA, USA}

\author{Christopher Russell}
\affiliation{Department of Earth, Planetary, and Space Sciences and Institute of Geophysics and Planetary Physics, \\University of California, Los Angeles, California, USA}

\author{Yoshizumi Miyoshi}
\affiliation{ISEE, Nagoya University, Nagoya 464‑8601, Japan}

\author{Satoshi Kasahara}
\affiliation{Department of Earth and Planetary Science, School of Science, The University of Tokyo}

\author{Ayako Matsuoka}
\affiliation{Institute of Space and Astronautical Science, Japan Aerospace Exploration Agency, Sagamihara, Kanagawa, Japan}

\author{Shoichiro Yokota}
\affiliation{Osaka University, Machikaneyama-cho, Toyonaka 560-0043, Japan}

\author{Kunihiro Keika}
\affiliation{Department of Earth and Planetary Science, School of Science, The University of Tokyo}

\author{Tomoaki Hori}
\affiliation{ISEE, Nagoya University, Nagoya 464‑8601, Japan}

\author{Yoichi Kazama}
\affiliation{Academia Sinica Institute of Astronomy and Astrophysics, No.1, Sec. 
Roosevelt Rd, Taipei 10617, Taiwan}

\author{Shiang-Yu Wang}
\affiliation{Academia Sinica Institute of Astronomy and Astrophysics, No.1, Sec.  Roosevelt Rd, Taipei 10617, Taiwan}

\author{Iku Shinohara}
\affiliation{ISAS, JAXA, Sagamihara 252‑0222, Japan}

\author{Yasunobu Ogawa}
\affiliation{National Institute of Polar Research, Tokyo, Japan}
\begin{abstract}
Magnetic field-line reconnection is a universal plasma process responsible for the conversion of magnetic field energy to the plasma heating and charged particle acceleration. Solar flares and Earth’s magnetospheric substorms are two most investigated dynamical systems where magnetic reconnection is believed to be responsible for global magnetic field reconfiguration and energization of plasma populations. Such a reconfiguration includes formation of a long-living current systems connecting the primary energy release region and cold dense conductive plasma of photosphere/ionosphere. In both flares and substorms the evolution of this current system correlates with formation and dynamics of energetic particle fluxes. Our study is focused on this similarity between flares and substorms. Using a wide range of datasets available for flare and substorm investigations, we compare qualitatively dynamics of currents and energetic particle fluxes for one flare and one substorm. We showed that there is a clear correlation between energetic particle bursts (associated with energy release due to magnetic reconnection) and magnetic field reconfiguration/formation of current system. We then discuss how datasets of in-situ measurements in the  magnetospheric substorm can help in interpretation of datasets gathered for the solar flare.   
\end{abstract}


\section{Introduction}
Magnetic field energy release and associated acceleration and heating of charged particles are common properties of powerful plasma phenomena such as flares in the solar atmosphere and substorms of the Earth's magnetosphere. This similarity has been underlined almost from the beginning of flare investigation \citep[e.g.,][]{Syrovatskii81} and actively discussed up to nowadays \citep{Terasawa00:AdSR, Reeves08:solar, Birn17:apj, Oka18:ssr, Sitnov19}. The basic elements of flares and substorms include formation of spatially localized regions with strong electric current density (sometimes in a form of current sheets, see \citet{bookParker94, bookBirn&Priest07, Rappazzo&Parker13}), magnetic field-line reconnection \citet{bookPriestForbes00:reconnection, book:Gonzalez&Parker}, energization of charged particles \citep{Aschwanden02, Zharkova11SSR, Birn12:SSR, Oka18:ssr}, and their precipitation to dense photosphere/ionosphere \citep[see comparison in, e.g.,][]{Haerendel09,Haerendel12} that is associated with a wide spectral range of emissions. Despite this similarity of a general picture, the plasma parameters and spatial scales of solar  flares and Earth's magnetospheric substorms are significantly different. This complicates any direct comparisons, but still leave the room for search of phenomenological similarities between these two systems.
 
One of the key elements of the magnetospheric substorms is the formation of a new type of current system that connects strong cross-field currents of hot rarefied (collisionless) magnetotail and field-aligned currents of cold dense (resistive) ionosphere \citep[e.g.,][]{Ganushkina15, Kepko14:ssr}. Development and dynamics of this current system (so-called substorm current wedge) occur on a global time-scale of the substorm, whereas spatially localized particle energizations and precipitations may intensify on much smaller time-scales. A combination of multi-spacecraft in-situ measurements in the magnetotail \citep[e.g., see reviews by][]{Petrukovich15:ssr, Sitnov19}, ground-based measurements of magnetic fields and various emissions \citep[see, e.g.,][]{Keiling09:ssr, Nishimura20:ssr}, and empirical models \citep[e.g.,][]{Stephens19, Sitnov19:jgr, Andreeva&Tsyganenko19} generally describes most of the elements of this current system and relations of these elements to such processes as particle energization and precipitation. But can this information be somehow useful for investigation of the solar flares that are mostly probed with remote observations and various reconstruction techniques? This study aims to address this question by comparing of two events – one moderate magnetosphere substorm and one moderate (M-class) solar flare. We would focus on two aspects of flares/substorms: (i) dynamics of magnetic field and electric currents and (ii) their correlation with charged particle energization.

Such a comparison of currents and magnetic fields for flares and substorms has become possible only recently, with a new observational dataset of solar physics -- high-cadence (135 s) photospheric vector magnetograms by the \textit{Helioseismic and Magnetic Imager} (HMI) on board the \textit{Solar Dynamics Observatory} \citep{Centeno14, Hoeksema14, Scherrer12}. Having lower signal-to-noise ratio than 12 min averaged magnetograms, these high-cadence magnetograms still provide sufficiently accurate measurements for long ($>2$ min) and intense (magnetic field perturbations higher than $\approx 300$ G) events \citep{Sun17:flare, Sharykin20}.

We start with detailed description of available datasets in Sect. \ref{sec:data}. This description continues our introduction, because we mix technical details of measurement techniques and data with their comparison for flares and substorms. Then we provide materials and interpretations of the spacecraft and ground-based measurements for a selected substorm in Sect. \ref{sec:substorm}. Comparison of these datasets and datasets available for the flare investigation is provided in Sect. \ref{sec:flare}. Finally we discuss obtained results and general perspective of using of the magnetospheric concepts/data for investigation of the solar flares (see Sect. \ref{sec:discussion}).

\section{Datasets and methods for substorms and solar flares \label{sec:data}}
\begin{figure}
\centering
\includegraphics[width=0.6\textwidth]{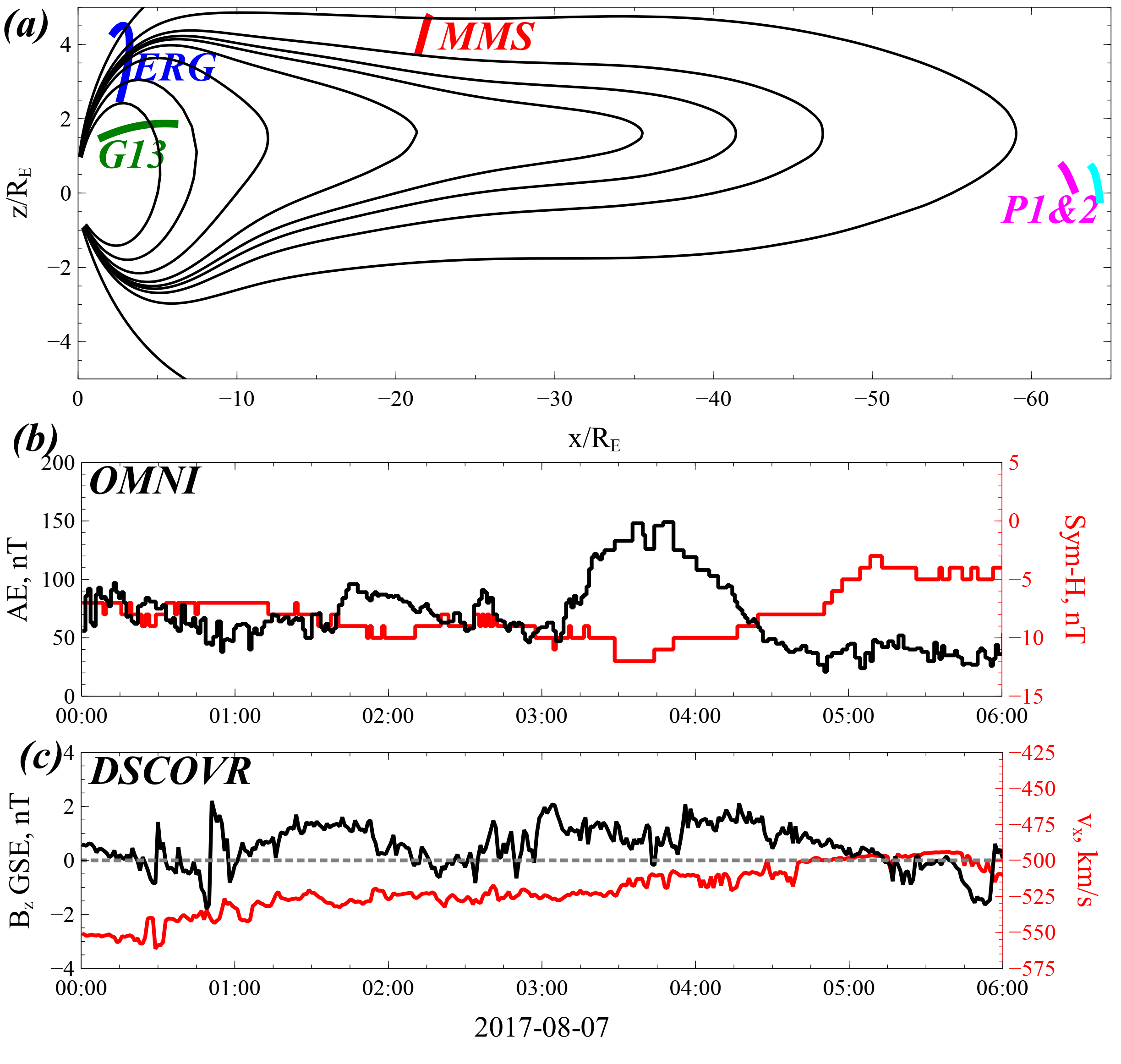}
\caption{\label{fig1s} Panel (a) shows spacecraft locations on top of magnetic field lines reconstructed from the empirical magnetic field model \citep{Tsyganenko02}. Panel (b) shows $AE$ and $Sym-H$ geomagnetic indexes. Panel (c) shows solar wind velocity $v_x$ and magnetic field $B_z$.}
\end{figure}

To describe available datasets of in-situ spacecraft measurements we show spacecraft locations relative to the magnetic field line configuration in the Earth's magnetotail (see Fig. \ref{fig1s}(a)). This configuration is from the empirical magnetic field model \citep{Tsyganenko02}, and we consider an interval before reconnection onset (before energy release). Two \textit{Acceleration, Reconnection, Turbulence and Electrodynamics of the Moon's Interaction with the Sun} (ARTEMIS) probes, $P_{1}$ and $P_{2}$ \citep{Angelopoulos11:ARTEMIS}, are located in the deep tail (around the moon; radial distance from the Earth is $r\sim60R_E$, $R_E$ is the Earth radius) and observe mainly outflow from the reconnection that is expected around $r\sim 30R_E$ \citep[statistically most probable region, see][]{Genestreti14}. In the context of solar flares these spacecraft may be considered as one measuring solar wind and mass ejection coming from the flare. ARTEMIS are equipped with flux-gate magnetometers \citep{Auster08:THEMIS}, electrostatic analyzers \citep[ESA,][]{McFadden08:THEMIS}, solid state detectors \citep[SST,][]{Angelopoulos08:sst}, and wave instruments \citep{Bonnell08,LeContel08,Cully08:ssr}. We will analyze magnetic field vector, ion and electron spectra ($<30$ keV from ESA and $50-500$ keV from SST, see \citet{Turner12:sst} for details of ESA-SST cross-callibration), and ion velocity vector provided in 4s (spacecraft spin) resolution. 

Closer to the Earth, in the magnetotail, four spacecraft of the \textit{Magnetospheric Multiscale} (MMS) mission \citep{Burch16} measured reconnection outflow moving toward the Earth (see  Fig. \ref{fig1s}(a)). Very small separation of these spacecraft makes their measurements almost identical to one another in our event (however, such small separation is extremely useful for investigation of a fine structure of the reconnection region, e.g. \citet{Burch16:science,Torbert18, Turner21:grl}). There is no direct analog of such earthward outflow measurements for the solar flares, but in more general sense MMS dataset would resemble estimates of outflow velocities and emissions from the energy release region in the solar corona which are obtained for some flares with the spatially and/or spectrally resolved observations in the soft X-ray and extreme ultraviolet (EUV) ranges  \citep[e.g.,][]{Hara11,Takasao12,Kumar13,Su13,Srivastava19}. We use four instruments onboard MMS: fluxgate magnetometer \citep{Russell16:mms}, fast plasma instrument \citep[FPI,][]{Pollock16:mms}, Energetic Ion Spectrometer \citep[EIS][]{Mauk16}, and Fly's Eye Energetic Particle Sensor \citep[FEEPS,][]{Blake16}. These instruments provide: magnetic field vector (64Hz are routinely available), ion and electron spectra ($<30$ form FPI, $50-500$ keV from EIS and FEEPS), and ion velocity vector with 3s (spacecraft spin) resolution.

At high latitudes (in the low-altitude magnetosphere), \textit{Exploration of energization and Radiation in Geospace} (ERG, Arase) spacecraft \citep{Miyoshi18:ERG} measured fluxes of energetic particles moving along magnetic field lines from the equatorial region (see  Fig. \ref{fig1s}(a)). This spacecraft formally probed the reconnection outflow, because particles quickly reach it along magnetic field lines. Despite its close location to the Earth, the local loss-cone size for ERG is still pretty small (several degrees) and most of observed energetic particles are trapped, i.e. will be reflected by the strong Earth magnetic field and move back to the equator. Again, there is no direct analog of ERG measurements to datasets available for solar flares, but these measurements may be considered as some sort of analog of emissions from the heated magnetic loop after energy release in the solar flare. In particular, the precipitation of accelerated electrons and protons into the dense layers of the solar atmosphere is observed as the sources of hard X-ray and gamma-rays, respectively, at the feet of the flare loops using, e.g., the \textit{Reuven Ramaty High Energy Solar Spectroscopic Imager} data \citep[RHESSI;][]{Lin02:rhessi,Hurford06,Krucker11}. Also, accelerated particles and conductive heat fluxes from the energy release site in the corona propagate down along the magnetic field and heat the plasma in the transition layer and the chromosphere, as a result of which it emits in the UV and optical ranges (e.g. in H$_{\alpha}$) -- the so-called flare bright kernels or flare ribbons are observed \citep[e.g., see the review by][]{Fletcher11}. We use four instruments on board ERG: fluxgate magnetometer \citep[MGF,][]{Matsuoka18:ERG_MGF, Matsuoka18:ERG_MGFdata},  electron Low-energy electron sensor \citep[LEPe][]{Kazama17:ERG_LEPe, Wang18:ERG_LEPedata}, ion and electron Medium-energy particle sensors \citep[MEPe and MEPi,][]{Kasahara18:ERG_MEPe, Kasahara18:ERG_MEPedata, Yokota17:ERG_MEPi, Yokota18:ERG_MEPidata}. These instruments provide: magnetic field vector, ion and electron spectra ($<20$keV from LEP-e, $10-180$ keV/q from MEP-i and $7-87$ keV from MEP-e) with 8-16s resolution.

At the equator, close to the Earth, the Geostationary Operational Environmental Satellite (GOES) spacecraft measured magnetic field perturbations. For sufficiently strong magnetosphere substorms GOES generally detect injections of energetic particles into the Earth's inner magnetosphere, but for the moderate event under consideration GOES only measures magnetic field perturbations \citep[with the fluxgate magnetometer, see][]{Singer96}. We use dataset with the time resolution of 1min. 

These four spacecraft missions (ARTEMIS, MMS, ERG, and GOES) provide in-situ magnetic field and plasma measurements for our event, whereas the spacecraft dataset is well supplemented by ground-based measurements. Magnetic field lines from the magnetotail for this event are projected to the north USA/Canada, where very large network of ground-based magnetometers are operating \citep{Mann08, Engebretson95, Russell08:ssr}. Such magnetometers provide magnetic field vector with 10s-1min resolution, and such fields show development of the ionospheric current system and ionosphere-magnetosphere currents \citep[e.g.,][]{Ganushkina15, Kepko14:ssr}. Owing to the good coverage of ground-based magnetometers, these measurements can be combined into the general dataset for reconstruction of currents on the ionosphere level \citep[][]{Amm&Viljanen99}. We use database of such currents provided by \citet{Weygand11,Weygand12} and containing transverse (to the Earth surface) current densities and field-aligned ($\sim$ normal to the surface) current with the time resolution of 10s. This dataset is the direct analog of currents and magnetic field on the photosphere level reconstructed for the solar flares \citep[e.g.,][]{Janvier14,Musset15,Sharykin20, Zimovets20}. Integral characteristics of ground-based magnetic field perturbations represent so-called geomagnetic indexes, e.g. $Sym-H$ and $AE$. The $Sym-H$ index, that is essentially the same as the $Dst$ index, quantifies longitudinally symmetric geomagnetic disturbances at mid-latitudes and serves as a measure of intensity of currents carried by high-energy ions injected deep into the inner magnetosphere (i.e., earthward from the GOES location in Fig. \ref{fig1s}). Such currents are going around the Earth and called ring currents \citep{Daglis99}. Essentially, $Sym-H$ describes the power of geomagnetic storm \citep[e.g.,][]{Gonzalez94}, much more intense perturbations than a substrom under consideration. Figure \ref{fig1s}(b) shows that $Sym-H$ index is quite small and there is no energetic ion injections penetrating sufficiently close to start drifting around the Earth for our event. However, $ASym-H$ index (that characterizes asymmetric magnetic field perturbations due to energetic ion enhancements at the night side) increased from $3$ to $6$ nT for our event (not shown). This is quite significant change indicating on energetic ion injection at the night side, where such injections can feed partial ring current \citep{Liemohn16, Sitnov19:jgr}. The Auroral Electrojet index, $AE$, measures auroral zone (where most of hot electrons from the magnetotail are precipitating, i.e. where a new current system connects magnetosphere and ionosphere currents) magnetic field perturbations and quantifies the substorm power \citep[e.g.,][]{Kamide&Rostoker04}. Figure \ref{fig1s}(b) shows $AE$ increase starting from $\sim$03:00 when the substorm onset has been observed by spacecraft in the magnetosphere (see below). Note we show AE averaged over all aurora zone stations around the Earth (the Kyoto AE), whereas AE calculated for the subset of stations projected to the substorm location is even larger and reaches $\sim 300$ nT in the peak at $\sim$03:30. Such AE index magnitude corresponds to the moderate substorm.

Ground-based magnetometer measurements are supplemented by all-sky imager measurements at South Pole \citep{Ogawa20}. The imager was in the premidnight sector (23.5 MLT at 3 UT, -74 deg MLAT), in the same sector as the satellites in the magnetosphere. The camera provides sufficiently high  spatial ($\sim 1$ km) and temporal resolution ($1$ and $4$ s at 557.7 and 630.0 nm wavelength) to detect emissions related to electron precipitations from the substorm magnetotail \citep{Chu&Nishimura20}. The emission heights at 557.7 and 630.0 nm are set to be 110 and 230 km altitude. Auroral emissions at 557.7 and 630.0 nm are sensitive to precipitating electrons above and below $\sim$ keV, respectively, which cover a similar energy range to soft X-ray emissions. 

In addition to ground-based magnetometers and all-sky cameras, this study makes use of data ($1$s resolution) from a 30 MHz riometer located in Iqaluit (IQA), Canada. This riometer is a passive sensor employing a single zenith-pointed antenna and measures the radio noise intensity from extra-terrestrial sources to characterize the opacity of the ionosphere \citep{Browne95, Danskin08}. Deviation of the received signal voltage from the voltage expected on a quiet day is used to derive signal absorption in decibels ($dB$). The riometer voltage is directly related to energetic electron precipitation to the cold dense ionosphere (riometers mainly respond to precipitations of energies above a few tens of keV; see, e.g., \citet{Gabrielse19}). Thus, the riometer dataset is the direct analogy of the bremsstrahlung hard X-ray emission produced by precipitating energetic electrons in solar flares \citep[e.g.,][]{Brown71, Syrovatskii&Shmeleva72, Kontar11}.

If geomagnetic indices monitor the dynamics of current systems in the magnetotail after reconnection onset, the drivers of this onset are generally associated with the solar wind that is monitored by spacecraft at L1 \citep{Acuna95, King&Papitashvili05, Koval13:agu}. Main solar wind characteristics affecting magnetotail dynamics and reconnection are plasma velocity (or dynamical pressure) jumps \citep[see, e.g.,][]{Angelopoulos20} and orientation of $B_z$ component of the solar wind magnetic field \citep[see, e.g.,][]{Baker96}. Figure \ref{fig1s}(c) shows that solar wind velocity is quite stable and $B_z$ is mostly positive: there are no conditions for strongly driven reconnection and the substorm under investigation is expected to be moderate.

\section{Currents and energetic particle fluxes}
We start with discussion of physical processes and their observational evidences in the magnetosphere substorm: from in-situ spacecraft observations to the ground-based magnetic field measurements. Then we present dataset of solar flare observations and comparison of substorm/flare datasets.

\subsection{Substorm \label{sec:substorm}}
\begin{figure}
\centering
\includegraphics[width=0.6\textwidth]{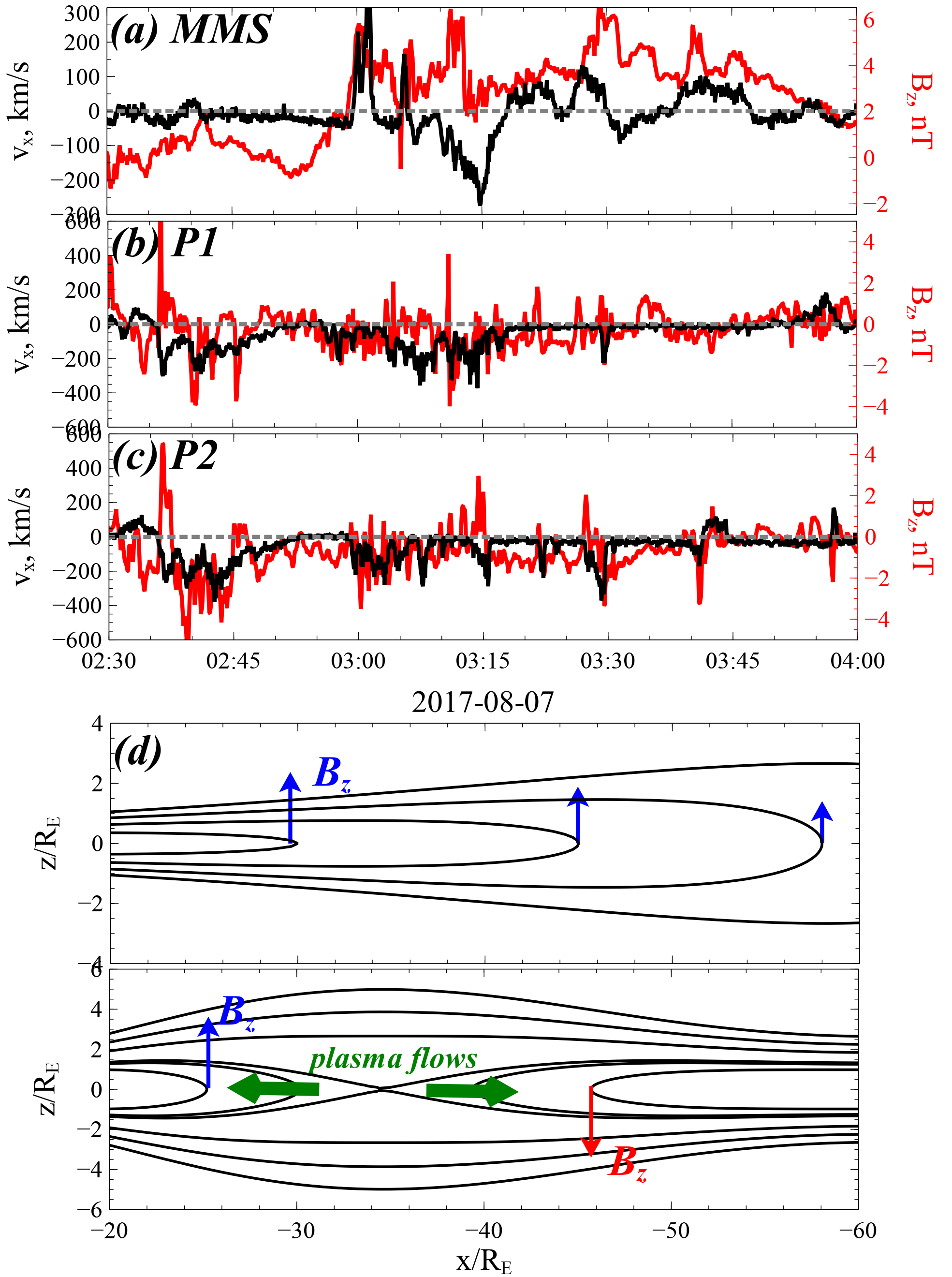}
\caption{\label{fig2s} MMS (a) and ARTEMIS (b,c) measurements of plasma flow velocity $v_x$ (black) and magnetic field $B_z$ (red). Panel (d) shows schematic of the magnetotail current sheet configuration before (top) and after (bottom) reconnection.}
\end{figure}

Figure \ref{fig2s}(a-c) shows ARTEMIS and MMS measurements of plasma flows along $x$ and $B_z$ magnetic field component (see Fig. \ref{fig2s}(d) for geometry of the reconnected current sheet in the Earth's magnetotail). MMS first observed $v_x>0$ and $B_z>0$ at $\sim$03:00: this is earthward plasma flow moving from the reconnection region and transporting magnetic flux (electric field $E_y\sim v_xB_z/c>0$). Both ARTEMIS probes at $\sim$03:00 observed $v_x<0$ and $B_z<0$: this is tailward plasma flow moving from the reconnection region and also transporting magnetic flux (electric field $E_y\sim v_xB_z/c>0$); see details in \cite{Angelopoulos13}. MMS and ARTEMIS observations suggest that the magnetic reconnection occurs right between these spacecraft, somewhere around $\sim-35R_E$ (see spacecraft locations in Fig. \ref{fig1s}).

ARTEMIS observations before 02:50 UT show tailward plasma flow ($v_x<0$, $B_z<0$) that indicates another reconnection occurring before the main one, at $\sim$03:00. However, during that time MMS spacecraft were quite far from the equatorial plane and did not see earthward flows ($v_x\sim 0$ on MMS), i.e. the energy release during this reconnection was not sufficiently strong to generate plasma flows that reconfigure the near-Earth magnetotail current sheet and expand plasma sheet to the MMS location. For the main event at $\sim$03:00 UT such flows destroy the near-Earth current sheet, increase its thickness (so-called plasma sheet expansion, see schematic in Fig. \ref{fig2s}(d) and \citet{Baker96, Baumjohann99,Baumjohann02}) and {\it move} MMS spacecraft to the equator where they observed $v_x>0$.

After $\sim$03:00 ARTEMIS observed bursts of tailward flows $v_x<0$, $B_z<0$ for almost forty minutes, whereas MMS observed oscillating $v_x$ and $B_z$. Transient nature of ARTEMIS observations is due to transient nature of reconnection that is not steady in the magnetotail \citep[e.g.,][]{Heyn&Semenov96, Semenov92,Sitnov09} and due to vertical oscillations of the magnetotail current sheet \citep[see][]{Runov05, Vasko15:jgr:CS} that periodically {\it move} spacecraft closer to or further from the equator. Oscillations of $v_x$ and $B_z$ on MMS are of more interesting nature and probably have some analogy in quasi-periodical vertical magnetic loop oscillations observed in some solar flares \citep[e.g.][]{Wang&Solanki04, Li&Gan06, Russell&Simoes&Fletcher15}. The earthward plasma flow $v_x>0$ transports magnetic flux to the region of strong dipole magnetic field, and impact of this flow compresses magnetic field lines there. This drives magnetic field lines and plasma oscillations along $x$ \citep[see][]{Panov10:grl, Panov13, Wolf18, Toffoletto20}, i.e. MMS may have observed quasi-periodical motions of magnetic field lines with $v_x>0$ periods associated with $B_z$ peaks (more dipolar configuration of magnetic field lines) and $v_x<0$ periods associated with $B_z>0$ minima (more stretched configurations of magnetic field lines).

\begin{figure}
\centering
\includegraphics[width=1\textwidth]{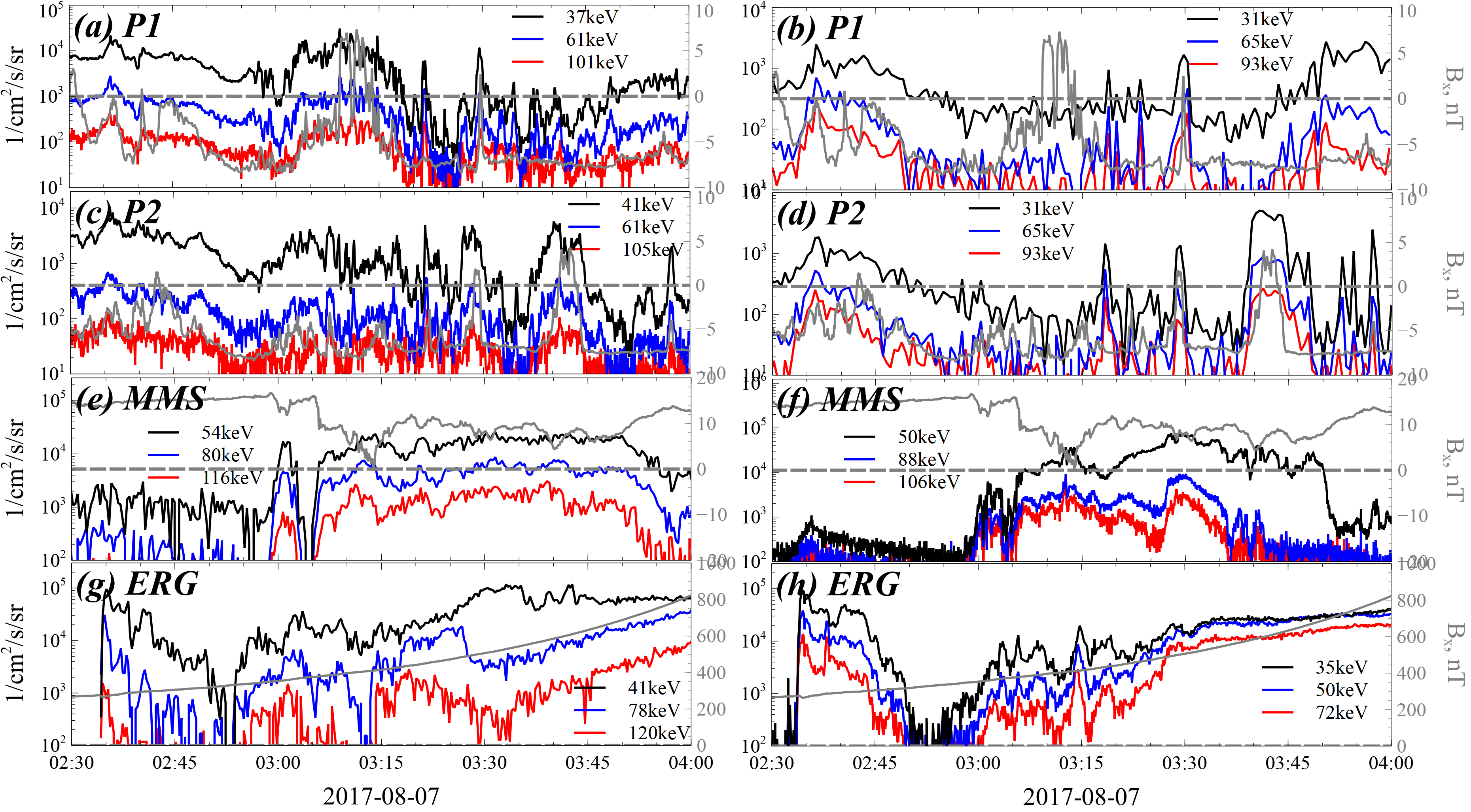}
\caption{\label{fig3s}ARTEMIS (a-d), MMS (e,f), and ERG (g,h) measurements of energetic ion fluxes (left) and energetic electron fluxes (right). Each panel also shows $B_x$ on the spacecraft (grey).}
\end{figure}

Figure \ref{fig3s} shows fluxes of energetic ions and electrons observed by MMS, ARTEMIS, and ERG. We also include $B_x$ profiles to separate temporal variations of energetic fluxes from their spatial variations: in the magnetotail $|B_x|$ is a proxy of the spacecraft location relative to the equator ($B_x=0$). At $\sim$03:00 MMS and ARTEMIS (mainly P1 that is closer to the equator than P2) observed rapid increase of energetic particle fluxes right after the reconnection. MMS are on the closed magnetic field lines, and energetic particles at the MMS side are trapped on these lines: they are bouncing between magnetic mirrors located near the Earth, where magnetic field well exceeds the magnetic field at the MMS location. This is one of the important differences between substorms and solar flares: loss-cone in the magnetotail is quite small and accelerated particles can be trapped on the closed magnetic field lines for a long time (e.g., MMS observed increased fluxes for 40 minutes after the primary energy release), whereas loss-cone in the solar flare configuration is much larger \citep{Aschwanden96, Aschwanden96:scaling} and a significant population of accelerated particles would precipitate right after energy release \citep[see discussion in][]{Eradat14}. Magnetic field-line shrinking (magnetic trap collapse), however, may reduce the precipitation and trap some accelerated particles in the top of flaring loops \citep[e.g.,][]{Karlicky&Barta06}. Additional trapping of particles from direct precipitations after acceleration may be due to pitch-angle scattering \citep[e.g.,][]{Kontar14, Charikov&Shabalin15} by strong turbulence \citep[that is also needed to explain magnetic field dissipation rate in the solar flares, see][]{Fleishman20}.  

Contrary to MMS, the ARTEMIS probes in the deep tail are on the open magnetic field lines. Energetic particles arriving to ARTEMIS from the reconnection region  quickly escape along open field lines to the solar wind. This explains why energetic particle fluxes on ARTEMIS are much smaller than fluxes at MMS \citep[see also][]{Runov18}. Note some fluctuations of fluxes on ARTEMIS are due to oscillations of the current sheet (see $B_x$ variations) that result in ARTEMIS motion away from the equator.

At $\sim$03:00, similarly to MMS, ERG observed increase of energetic particle fluxes. These energetic particles arrive at ERG along magnetic field lines, and these are trapped particles bouncing along magnetic field lines. MMS spacecraft before $\sim$03:00 were on magnetotail field lines, far from the equator (see large $B_x$), and can be projected to the deep tail where fluxes of energetic particles are very low (there is no MMS observation of $\sim 100$ keV fluxes before $\sim$03:00). ERG before $\sim$03:00 was in the inner magnetosphere and projected to the near-Earth region where energetic particle fluxes can be quite high, as particles of these energies are trapped there by strong magnetic field and can survive from the previous energy release. This explains why ERG observed sufficiently large fluxes even before $\sim$03:00. Dynamics of energetic fluxes on ERG and MMS can be compared with currents and magnetic field dynamics on the ground, to establish relations between formation of new current systems and energy release process. 

\begin{figure}
\centering
\includegraphics[width=0.75\textwidth]{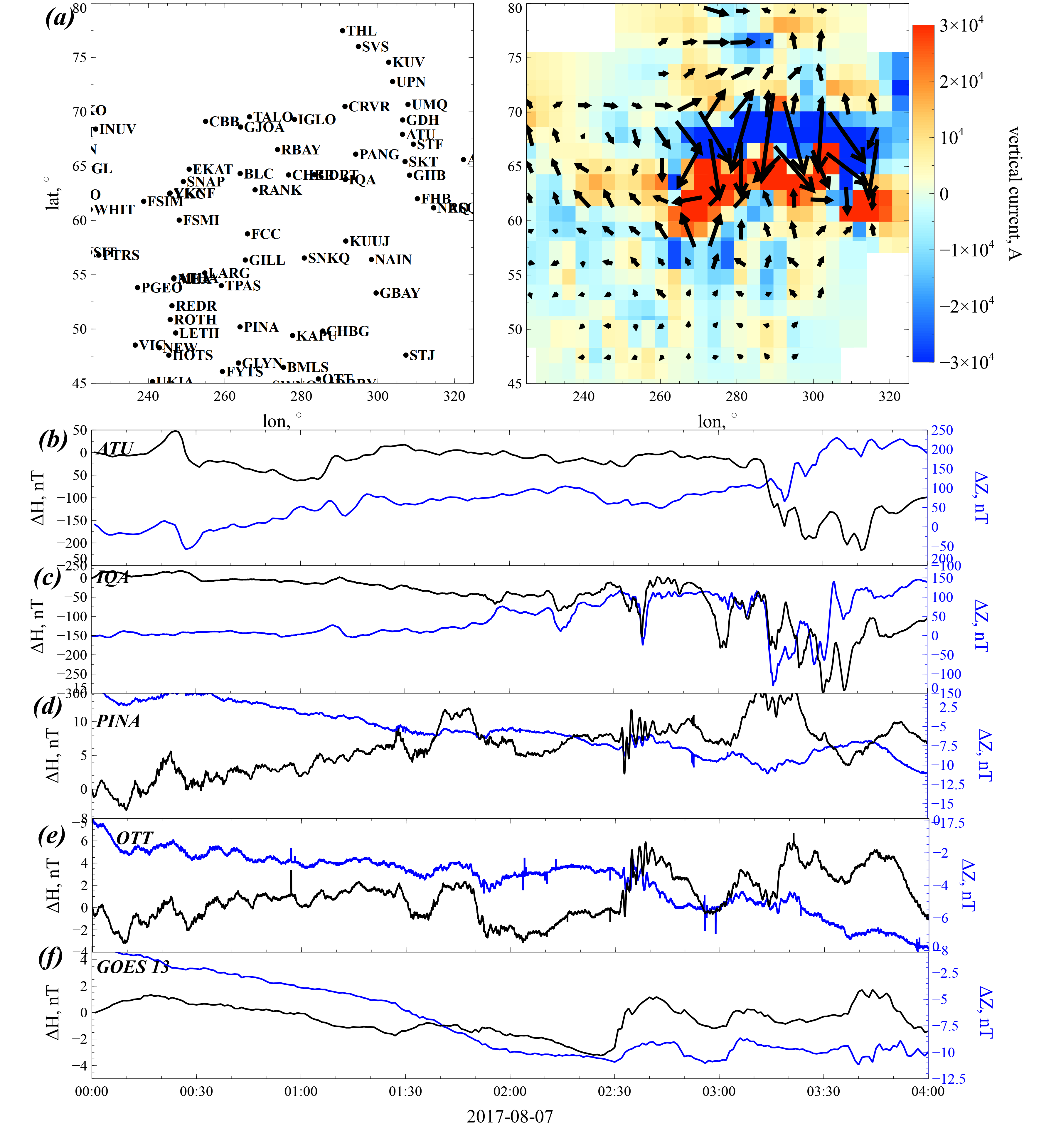}
\caption{\label{fig4s} Panel (a) shows location (in geographic latitude, longitude) of the ground-based stations measuring magnetic field (left), density of the transverse current (black arrows), and intensity of downward and upward vertical currents in red and blue at 03:20 UT (right). Panels (b-f) shows variations of horizontal $H$ and vertical $Z$ components of magnetic field on four ground-based stations and on the near-Earth GOES spacecraft.}
\end{figure}

Figure \ref{fig4s}(a) shows locations of ground-based magnetometer stations measuring magnetic field that can be used to calculate the horizontal (shown by black arrows) and vertical (color) currents. Note the main background magnetic field is much stronger than magnetic field perturbations and is directed nearly normal (vertical) to the ionosphere surface, and thus the horizontal and vertical currents are approximately perpendicular and field-aligned currents. There are two clear field-aligned current peaks connected by transverse current, i.e. current flows from the magnetotail along magnetic field lines, then flows transversely the magnetic field (along the ionosphere surface), and finally flows back to the magnetotail. This is the ionospheric part of the current wedge \cite{Kepko14:ssr} that is closed by the cross-field current in high $\beta$ magnetotail current sheet. 

Stations within the peaks of downward and upward currents detect decrease of the horizontal (parallel to surface) magnetic field (see $\Delta H<0$ in Fig. \ref{fig4s}(b,c), ATU and IQA), and increase (for inflow)/decrease (for outflow) vertical component of the magnetic field (see $\Delta Z>0$/$\Delta Z<0$  in Fig. \ref{fig4s}(b,c)). Note the orientation of field-aligned currents and corresponding magnetic field perturbations can vary from substorm to substorm or even within one substorm. The more typical substorm current wedge has downward field-aligned currents on the dawnside, upward field-aligned currents on the duskside, and westward current in between. It gives southward magnetic field perturbation ($\Delta H < 0$). At the time of Fig. \ref{fig4s}(a), the currents are mostly north-south, and thus the main perturbation of the magnetic field is east-west ($\Delta Z$). The upward and downward currents are separated north-south. 

Stations located below the main inflow/outflow current region are projected to the magnetotail equator closer to the Earth than the main energy release region and fast plasma flows. These stations detect increase of the horizontal magnetic field (see $\Delta H>0$ in Fig. \ref{fig4s}(d,e), PINA and OTT) that indicates the magnetic field increase in the near-Earth region due to transport there the magnetic flux by fast plasma flows \citep{Kokubun&McPherron81}. Indeed,  $\Delta H>0$ from Fig. \ref{fig4s}(d,e) correlates well with $\Delta H>0$ measured by GOES in the near-Earth magnetotail (see Fig. \ref{fig4s}(f)).

\begin{figure}
\centering
\includegraphics[width=0.6\textwidth]{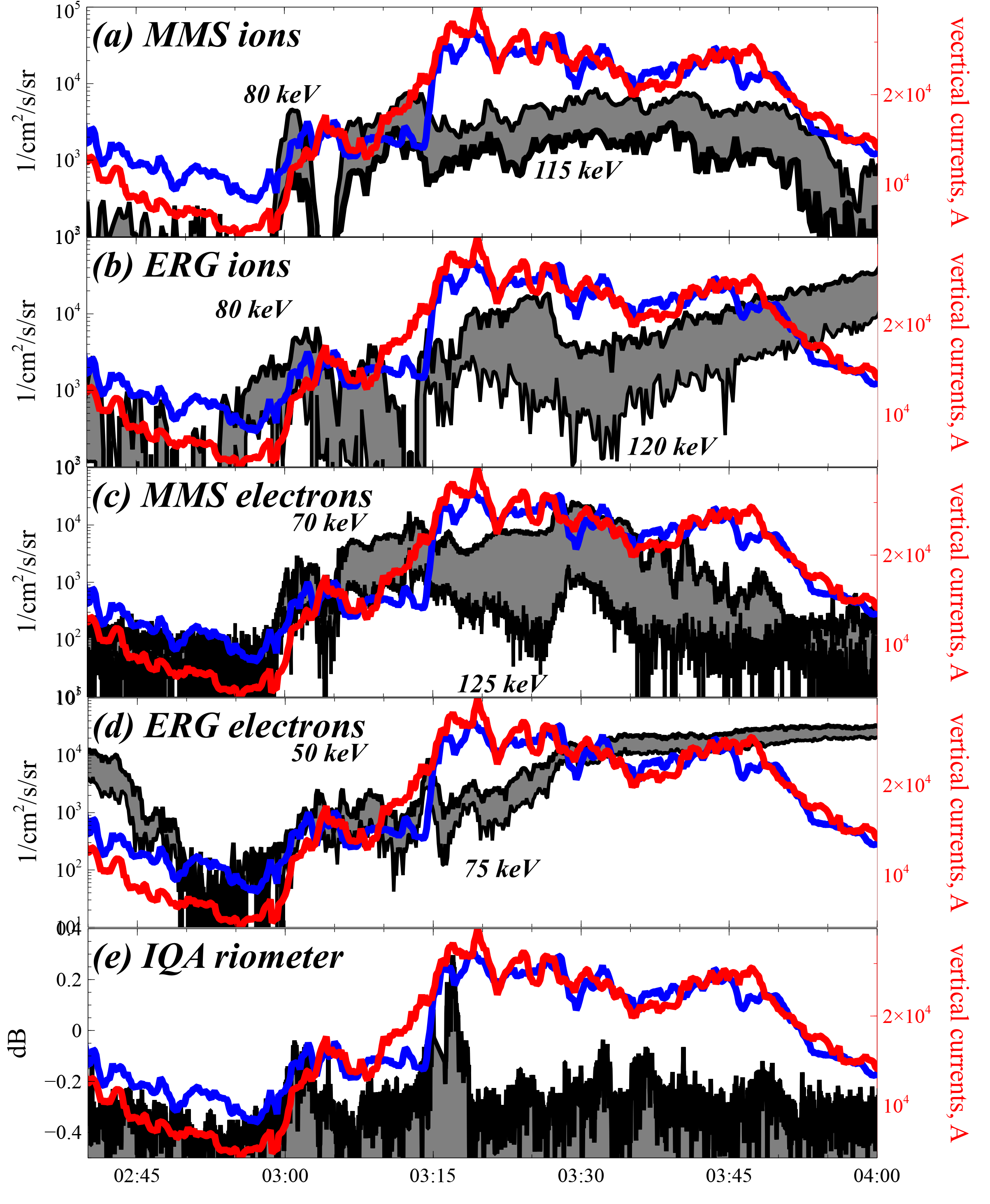}
\caption{\label{fig5s} Downward (red) and upward (blue) currents from Fig. \ref{fig4s} (averaged over the region with the current magnitude larger than $10^3$A) are shown together with energetic particle fluxes and precipitation signals:  (a\&c) electrons and ions measured by MMS, (b\&d) electrons and ions measured by ERG, (e) riometer data from IQA station.}
\end{figure}

Figure \ref{fig5s} compares dynamics of currents and energetic particle fluxes. We consider field-aligned (inflow and outflow) currents averaged over the regions with current magnitude larger than $10^3$A. To trace dynamics of energetic particle fluxes in the magnetotail we show ion and electron fluxes from ERG and MMS, whereas the enhancement of the riometer signal (reduction of signal voltage) indicate enhancements of precipitating electron fluxes (the riometer station is located at IQA, see Fig.\ref{fig4s}(a)). There are two main growths of the current magnitude: at the main energy release, $\sim$03:00, and at $\sim$03:15. The second growth is likely due to another energy release that is not seen by MMS, but can be distinguished in ERG fluxes (see Fig. \ref{fig3s}). Both current growth intervals are associated with strong precipitations of energetic electrons detected by riometers (see Fig. \ref{fig5s}(e)). Such temporally localized precipitations are due to combined effect of energy transport to the ionosphere by Alfven waves \citep{Lysak04, Keiling09:ssr} generated at the energy release region (and following dissipation of these waves with electron field-aligned acceleration in the ionosphere/magnetosphere interface; see \citet{Lysak&Song03, Chaston05, Chaston08, Malaspina15:kaw}) and direct precipitations of energetic electrons scattered around equator to the loss-cone by various transient electromagnetic fields generated at the energy release region \citep{Eshetu18, Gabrielse19}.   

Energetic ion and electron fluxes at MMS do not show two peaks, but rather repeat the profiles of the currents (see Fig. \ref{fig5s}(a,c)). There is a clear increase of fluxes due to electron and ion acceleration during the energy release (magnetic reconnection) at $\sim$03:00. Then fluxes stay on the stable level for $\sim$40 minutes and decay together with the currents around $\sim$03:45. Similar flux dynamics is shown by ERG observations (see Fig. \ref{fig5s}(b,d)), but ERG also observed the second peak of energetic fluxes around $\sim$03:15, when currents increase. After $\sim$03:30 ERG, moving along highly inclined orbit, lost the conjugation to the magnetotail and started observing fluxes on the inner magnetosphere that are not related to the current dynamics.

Comparison of MMS and riometer data (see Fig.\ref{fig5s}(a,e)) shows that precipitations of energetic electrons correlate well with the intervals of current increase, whereas the current slow evolution is better correlated with magnitude of equatorial fluxes. Indeed, currents derived from the ground-based magnetometer measurements are connected to the magnetotail cross-field currents supported by drifts of energetic particles seen by MMS. Thus, increases of current magnitude should be due to increase of equatorial energetic particle fluxes that are associated with energy releases and precipitations. In the next section we discuss the analogy of this scenario with observations of emissions and currents in the solar flare.

\begin{figure}
\centering
\includegraphics[width=1\textwidth]{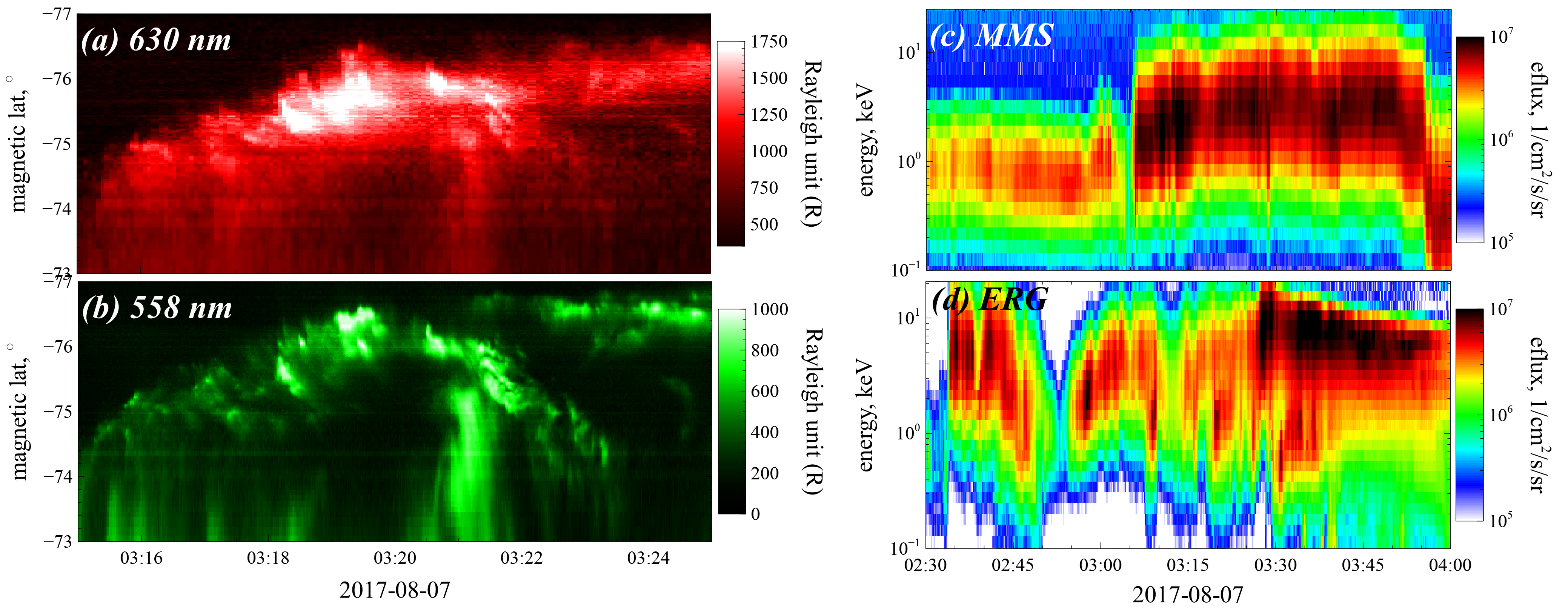}
\caption{Panels (a) and (b) show Keograms (bright emission from low latitudes corresponding to the near-Earth region) for red (corresponding to precipitations of $\lesssim 1$ keV electrons) and green (corresponding to precipitations of $\gtrsim 1$ keV electrons) spectral lines (the color scale is Rayleigh unit (R) with $1$ R = $10^10 $photon$/$cm$^2/$s, see \citet{Ogawa20:norm}). The magnetic latitudes are in the AACGM (altitude adjusted corrected geomagnetic) coordinate system. Panels (c) and (d) show electron spectra measured by MMS and ERG spacecraft. \label{fig6s}}
\end{figure}

Ground-based measurements are not limited to magnetic fields and riometer signals, but also include optical measurements by an all-sky-camera at South Pole. This dataset represent 2D time dependent pictures of emissions at different wavelength, corresponding to precipitations of electrons of different energies. We consider red (630.0 nm) and green (557.7 nm) light emissions of $\leq 1$ keV and $\geq 1$ keV electron precipitations ($\geq1$ keV can be considered as a suprathermal part of the main plasma population in the magnetotail that are enhanced and scattered into the loss-cone during energy release, whereas $\leq 1$ keV is the main plasma sheet and lowest energy part of energetic population; see discussion in \citet{Runov15,Ganushkina19}).  Figure \ref{fig6s}(a,b) shows keograms, the intensity of emissions as function of time and latitude at a fixed longitude. We supplement keograms by thermal electron spectra measured by MMS and ERG (see Fig. \ref{fig6s}(c,d)). These spectra show clear heating of the main electron population (flux increase) at $\sim$03:00 (note ERG captured another flux increase at $\sim$03:15 in agreement with two bursts of electron precipitations shown by riometers, see Fig. \ref{fig5s}(a)).

Keograms show bright emission from low latitudes corresponding to the near-Earth region where plasma flows transport energetic and thermal fluxes. Then emission gradually expand to higher latitudes, i.e. the near-Earth region filled by high fluxes of energetic and thermal particles expands back to the tail. Interesting feature of red/green light comparison is that precipitations of $\leq 1$ keV electrons are much more {\it diffusive} and limited to the poleward portion, whereas precipitations of $\geq 1$ keV electrons are more structured and distribute over a wider latitude range, and at each moment of time they consist of several bright bursts. While the diffuse nature is partly due to the slower de-excitation time scale of the 630.0 nm emission, such observations can also be interpreted in a context of difference between large-scale heating of the magnetotail thermal electron population and spatially localized accelerations of energetic electrons. Indeed, heating can be provided by adiabatic mechanisms driven by enhanced plasma convection to the Earth (charged particle transport to the higher magnetic field region), whereas acceleration to high energies requires sufficiently intense electric fields that are typically localized around magnetic reconnection \citep{Nagai13:statistics,Nagai13, Torbert18}, leading fronts of fast plasma flows \citep{Runov09grl, Runov11jgr}, and right around the aurora acceleration region \citep[e.g.,][]{Partamies08, Birn12:SSR, Haerendel21}. The red emission at higher latitudes corresponds to lower-energy electrons at larger distance from the Earth, while the dominance of the green emission at lower latitudes corresponds to electron energies increasing to lower distance from the Earth as electrons are transported earthward. There is an  analogy to large-scale heating of the plasma within the entire flare loop (resulting in enhanced soft X-ray and EUV emissions and increase of the plasma temperature without formation of nonthermal populations, see \citet{Fleishman15}) and much more localized bursts of hard X-ray emission associated with the energetic electrons (see more discussion in the next section). In both systems (magnetotail and solar flares) the direct acceleration within the reconnection region is not very effective \citep[e.g.,][]{Imada11, Turner21:grl, Cohen21:grl}  due to the localization of this region and instability of particle motion there \citep[see discussion in][]{Bulanov76, Zelenyi90:acceleration, Vekstein&Browning97, Browning&Vekstein01, Litvinenko96, Litvinenko03, Hannah&Fletcher06, Artemyev14:angeo}. However, reconnection releases a significant portion of energy in a form of plasma flows, that collapse magnetic field and effectively accelerate large plasma volumes in the magnetotail \citep[e.g.,][]{Imada07, Birn14, Runov15, Sorathia18} and solar flares \citep[e.g.,][]{Karlicky&Barta06, Bogachev07, Eradat&Neukirch14, Borissov16}.

\subsection{Solar Flare \label{sec:flare}}


To compare with the magnetospheric substorm analysis, we consider a moderate M6.5 long-duration eruptive solar flare happened in the active region 12371 near the solar disk center (around N13W06) on 22 June 2015. The flare was accompanied by a coronal mass ejection (CME) with a projected speed of $\approx 1200$ km/s \citep{Vemareddy17ApJ}. Various aspects of this flare have already been studied and discussed in more than a dozen works \citep[e.g.][]{Jing16NatSR, Liu16NatCom, Bi17ApJ, Vemareddy17ApJ, Wang17NatAs, Kuroda18ApJ, Liu18ApJ, Wang18ApJ, Wheatland18ApJ, Xu18NatCo, Kang19ApJ}. The flare attracted widespread attention, first of all, as it was very well observed by a number of modern instruments. In particular, the flare region was observed in details in the optical range with very high spatial resolution (up to $\approx 70$ km) with the 1.6-m Goode Solar Telescope \citep[GST; formerly known as the New Solar Telescope, NST; e.g.][]{Goode12SPIE} at Big Bear Solar Observatory (BBSO). Another important factor, which we exploit in the present work, is that vector photospheric magnetograms made with the Helioseismic and Magnetic Imager \citep[HMI;][]{Scherrer12,Hoeksema14} onboard the Solar Dynamics Observatory \citep[SDO;][]{Pesnell12SoPh} with a small time step of 135 s are available for the time interval of this flare \citep{Sun17ApJ}. This makes it possible to study the dynamics of the magnetic field vector and vertical electric current at the photosphere during the flare. The flare has also been observed by several other ground-based and space-based observatories providing information on the development of flare sources in various spectral ranges. In particular, we use the images of the Sun made in the EUV and UV emissions by the Atmospheric Imaging Assembly \citep[SDO;][]{Lemen12SoPh} onboard SDO, and in the X-ray range by RHESSI.

\subsubsection{Flare Emission Sources \label{sec:flaresources}}

\begin{figure}
\centering
\includegraphics[width=0.7\textwidth]{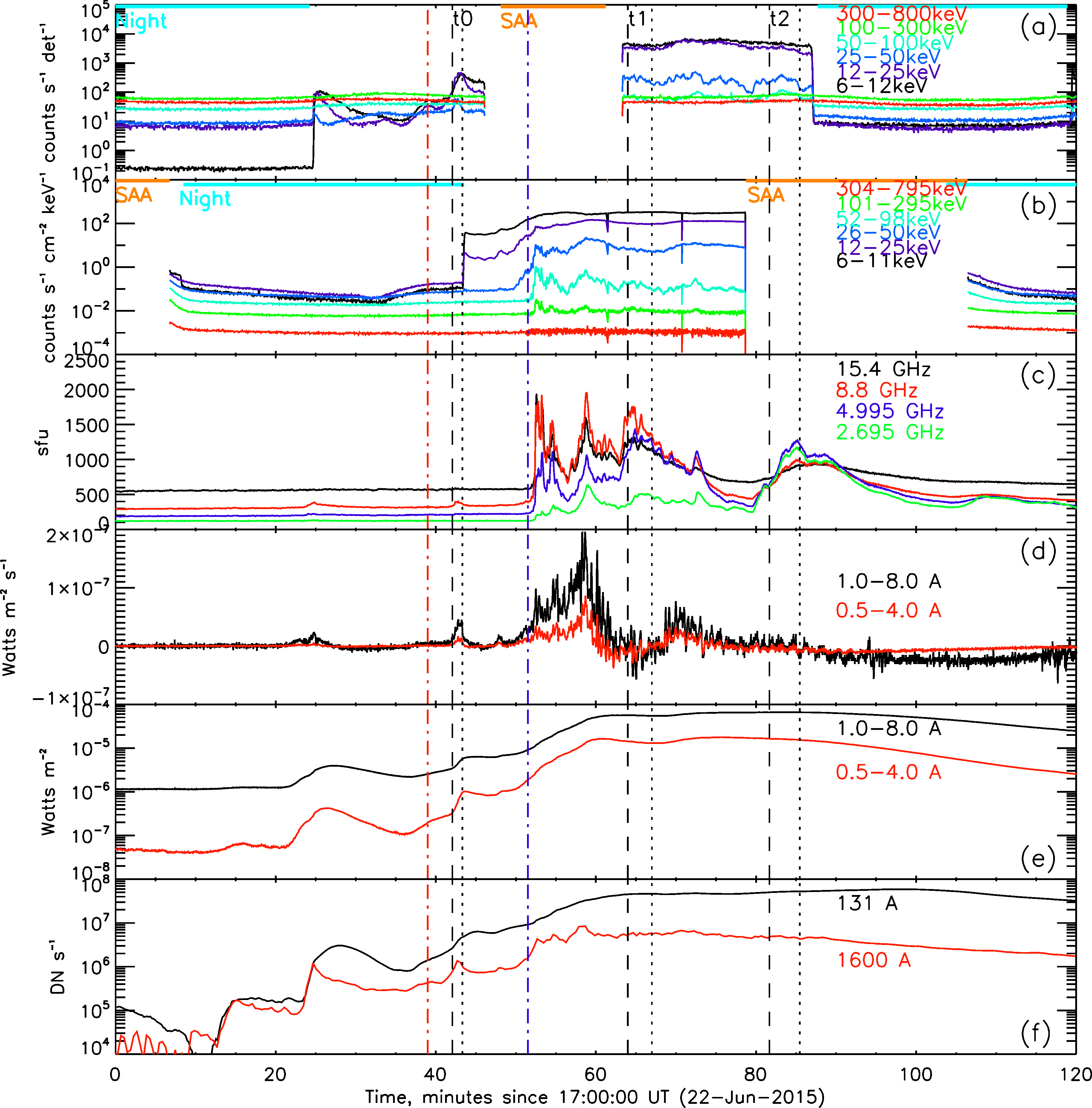}
\caption{\label{fig1f} Time profiles of the M6.5 solar flare electromagnetic emissions on 2015 June 22. (a) Corrected RHESSI count rates in the six standard X-ray energy channels. The time intervals when the spacecraft was in the shadow of the Earth (Night) and in the South Atlantic Anomaly (SAA) are shown above with the cyan and orange horizontal bold lines, respectively. (b) Fermi/GBM count rates in the six similar X-ray energy bands. The Night and SAA time intervals for the Fermi spacecraft are also shown on the top with the same colors as in (a). (c) Flux density of solar microwave emission at four fixed frequencies detected by one of the Radio Solar Telescope Network (RSTN) observatory in Palehua. (d) Time derivative of the GOES/XRS $1.0–8.0$ \AA{} (black) and $0.5-1.0$ \AA{} (red) X-ray light curves, which are shown on (e) by the same colors. (f) The pre-flare background-subtracted light curves of the EUV and UV emissions in the SDO/AIA 131 \AA{} and 1600 \AA{} channels integrated over the flare region which is shown in Fig. \ref{fig2f}. The vertical dash-dotted red and blue lines indicate the {\it official} flare start, at 17:39 UT, according to the GOES observations in the $1.0–8.0$ \AA{} channel and the begin of the flare impulsive phase (at around 17:51:30 UT), respectively. The vertical black dashed and dotted lines show the begin and end, respectively, of three time intervals ($t_{0}$, $t_{1}$, and $t_{2}$) for which the images of the flare X-ray sources shown in Fig. \ref{fig2f} are constructed.}
\end{figure}

Fig. \ref{fig1f} gives an overview of temporal evolution of electromagnetic emission of this flare in different spectral ranges. The {\it official} start of the flare was at 17:39 UT according to the observations in the X-Ray Sensor \citep[XRS;][]{Garcia94SoPh} 1-8 \AA{} soft X-ray channel onboard the Geostationary Operational Environmental Satellite (GOES), and the peak time was at 18:23 UT. The flare lasted more than 3 h in the soft X-ray range. The flare onset was preceded by several episodes of energy release (precursors), which were investigated in details by \citet{Wang17NatAs}. It was shown that these small-scale episodes of energy release happened in a small magnetic channel near the polarity inversion line (PIL) in the footpoints of highly-sheared loops, in which the flare developed later. The precursors were accompanied by an increase and decrease in the magnetic flux and vertical electric current. \citet{Wang17NatAs} suggested that the precursors were accompanied by emergence of a small current-carrying flux tube, which might be dissipated by reconnection with surrounding field during or after the flare. Based on the analysis of 3D magnetic structure reconstructed in the nonlinear force-free field (NLFFF) approximation, \citet{Kang19ApJ} proposed a three-step scenario for the initiation of eruption leading to this flare: (1) the formation of the double arc loops by the sheared arcade loops through the tether-cutting reconnection in the early flare phase (in the precursors), (2) expansion of the destabilized double arc loops due to the double arc instability (DAI), and (3) the full eruption due to the torus instability leading to the flare and various accompanying secondary phenomena discussed in the aforementioned works and in the present paper.

The impulsive phase of the flare began at about 17:51:30 UT and was accompanied by the appearance of a sequence of bursts of nonthermal hard X-ray and microwave radiation associated with the appearance of populations of accelerated electrons in flare loops. Unfortunately, RHESSI stayed in the South Atlantic Anomaly (SAA) approximately in the time interval from 17:46 UT to 18:03 UT, and for this reason missed the beginning of the impulsive phase (Fig. \ref{fig1f}(a)). However, hard X-ray emission from the flare during this time interval was well detected by the Fermi Gamma-Ray Burst Monitor \citep[GBM;][ see (Fig. \ref{fig1f}(b))]{Meegan09ApJ}. Time derivative of the soft X-ray emission measured by GOES/XRS, roughly mimics the nonthermal flare emissions, at least in the first half of the impulsive phase \citep[the so-called Neupert effect;][]{Neupert68,Dennis93SoPh}, and has the main peak around 17:58:36 UT (Fig. \ref{fig1f}(d)). At about the same time, a peak of UV radiation was observed in the SDO/AIA 1600 \AA{} channel, emitted from the flare ribbons in the chromospheric feet of the flare loops. This peak roughly coincides with the peaks of hard X-ray and microwave radiation (Fig. \ref{fig1f}(b,c)), indicating that, at least at this time, accelerated electrons precipitating from coronal sources could efficiently heat the chromospheric plasma.

\begin{figure}
\centering
\includegraphics[width=0.8\textwidth]{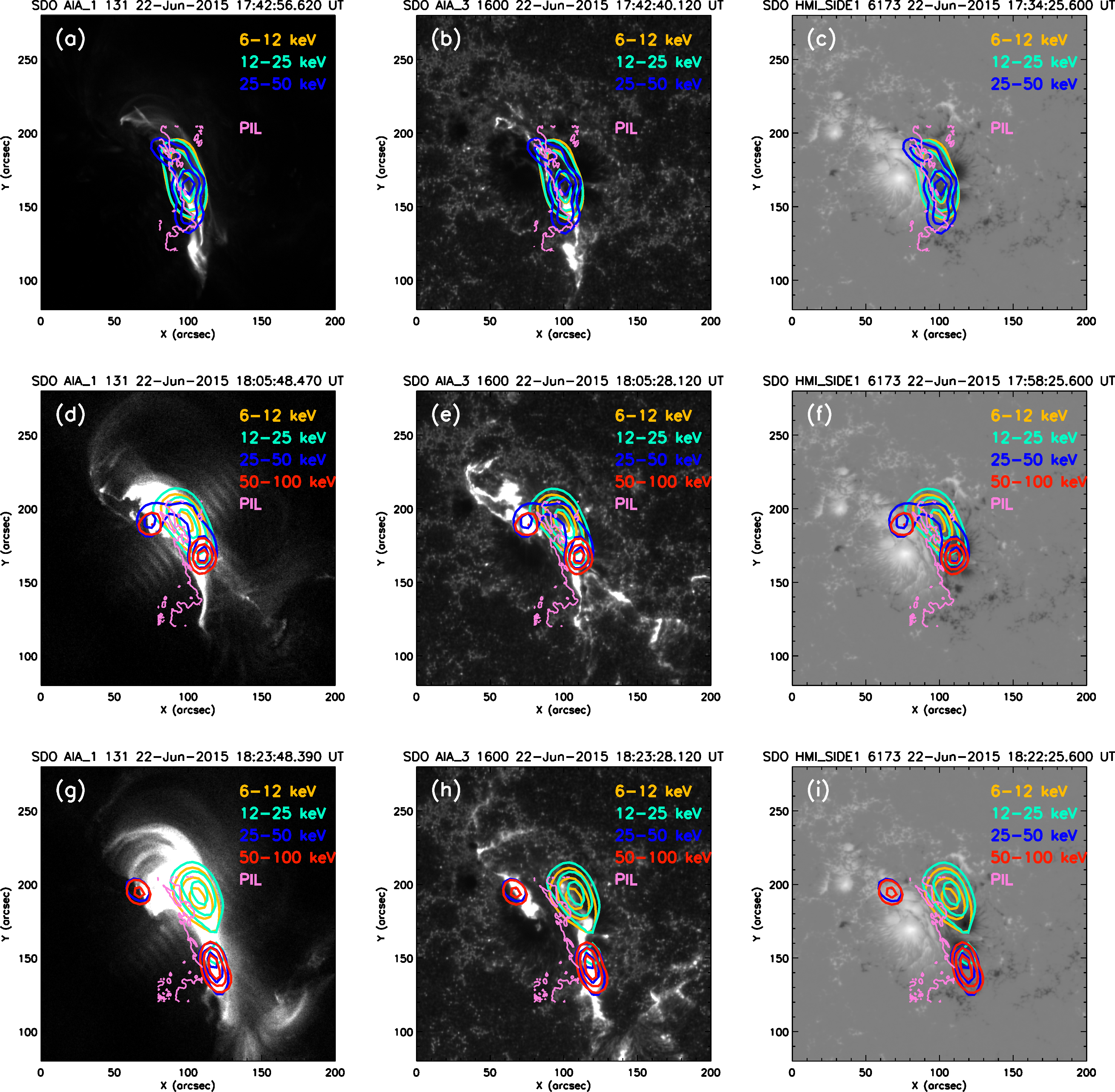}
\caption{\label{fig2f} Images of the studied solar flare region for the three time intervals $t_{0}$ (a--c), $t_{1}$ (d--f), and $t_{2}$ (g--i) indicated by the vertical dashed and dotted black lines in Figure \ref{fig1f}. The background images in the left column show flare loops in the corona observed in the SDO/AIA 131 \AA{} channel. The background images in the middle column show flare ribbons in the chromosphere and transition region observed in the SDO/AIA 1600 \AA{} channel. The background images in the right column are the 720 s line-of-sight photospheric magnetograms made with the SDO/HMI instrument data. The color palette is ranging from $-2500$ G (black) to $+2500$ G (white). The observational times are shown above the figures. The curves of different colors show isocontours at a level of 50\% of the maximum brightness of X-ray sources in the ranges $6-12$ (orange), $12-25$ (cyan), $25-50$ (blue) and $50-100$ (red) keV, reconstructed in the time intervals $t_{0}$, $t_{1}$ and $t_{2}$ from the RHESSI data using the \textit{Clean} algorithm. The pink contours show the magnetic polarity inversion line (PIL) at the photosphere in the central part of the flare region.}
\end{figure}

As an example, in Fig. \ref{fig2f}, we present images of flare sources of soft ($<10-20$ keV) and hard X-ray emission for three time intervals: $t_{0}$ (17:42:40 -- 17:43:20 UT) at the beginning of the impulsive phase (panels (a-c)), $t_{1}$ (18:04:00 -- 18:07:00 UT) including several emission peaks in the impulsive phase (panels (d-f)), and $t_{2}$ (18:21:40 -- 18:25:28 UT) in the vicinity of the main peak of the flare soft X-ray emission (panels (g-i)). The start and end times of these intervals are shown in Fig. \ref{fig1f} with the black vertical dashed and dotted lines, respectively. The duration of the selected intervals is rather long in order to provide a sufficient signal-to-noise ratio for constructing high-quality X-ray images. X-ray images were constructed from the observation data of RHESSI front detectors 3, 4, 5, 6, 7, and 8 using the Clean algorithm \citep{Hurford02SoPh}. The images of X-ray sources are superimposed on the images made with SDO/AIA in the 131 \AA{} (panels (a, d, g)) and 1600 \AA{} (panels (b, e, h)) channels, and on the nearby 720 s HMI/SDO magnetograms of the line-of-sight magnetic field component (panels (c, f, i)), which is close to the radial magnetic component, $B_{r}$, because of near the solar disk center location of the flare region. Images in the 131 \AA{} channel predominantly show the flare loops filled with hot plasma heated to temperatures $T \sim 10$ MK ($\sim 1$ keV), while images in the 1600 \AA{} channel show the flare ribbons in the chromosphere and transition region \citep[more details about the flare ribbons in this event were presented by][]{Jing16NatSR, Liu18ApJ}. All images also show the position of the magnetic PIL on the photosphere (pink curves), as one of the traditional landmarks of the flare region. It can be seen that the impulsive phase of the flare (around $t_{0})$ began in a system of highly sheared loops extended along the PIL \citep[see also][]{Wang17NatAs, Kang19ApJ}. At this time, X-ray sources were co-located with these loops above the PIL. No hard X-ray sources with energies above around 50 keV were visible. The flare ribbons were very close to the PIL at that time. As the flare progressed (see Fig. \ref{fig2f}(d-f) and (g-i)), the ribbons grew larger and moved away from the PIL, the flare loops became larger and taller indicating an increase in the height of the primary energy release cite(s) in the corona with time. One can see appearance of hard X-ray footpoints with energies $\sim 100$ keV in $t_{1}$ and $t_{2}$ indicating the precipitation of accelerated electrons of corresponding energies into the feet of the flare loops in the chromosphere in those time intervals.


Hard X-ray and microwave emissions in the flare impulsive phase, near 18:05:32 UT (within $t_{1}$), were investigated by \citet{Kuroda18ApJ} using the spatially and spectrally resolved observations by RHESSI and the Expanded Owens Valley Solar Array (EOWSA), respectively. Combining the NLFFF extrapolation and 3D modelling within the GX Simulator \citep[e.g][]{Nita15ApJ} it was found that there could be at least two populations of accelerated electrons in the flare region. The first one with a broken power-law spectrum ($E_{break} \approx 180-220$ keV) was trapped in the flare loops producing the high-frequency part of the microwave spectrum, and the precipitating part of these electrons produced the observed hard X-ray footpoints. The second population of non-thermal electrons was trapped in a large volume (called as the {\it overarching loop}) with relatively weak magnetic field above the main flare loops and was invisible in the hard X-ray range according to the RHESSI observations, while it was observed by its gyrosynchrotron radiation emitted mainly in the low-frequency microwave range. In overall, the results by \citet{Kuroda18ApJ} showed good correspondence with the {\it standard} 3D eruptive flare model, where electrons are accelerated in the reconnection cite(s) in the corona, and trapped and precipitated in the complicated system of multiple underlying loops.

UV (and optical) emission from the flare ribbons show heating of plasma in the feet of flare loops connected to the energy release cite(s) in the corona. The similar heating is observed in the magnetotail current sheet during the substorm and detected both by in-situ measurements and optical observations of electron precipitations into the ionosphere (see Fig. \ref{fig6s}). However, in contrast to collisional heating within flare loops, the main plasma heating mechanism in the magnetosphere is the collisionless adiabatic heating of electrons and ions trapped within shrinking magnetic flux tubes \citep{Birn15, Ukhorskiy18:DF, Eshetu19}. This heating mechanism resembles the magnetic trap collapse described in application to solar flares by \citet{Karlicky&Kosugi04, Bogachev05, Bogachev07, Borissov16}. Such heating at the top of the corona loop can be quite effective both in 2D and 3D magnetic trap configurations \citep[see estimates in][]{Grady&Neukirch09}, but the main plasma heating within the loops (that one responsible for the UV and optical emissions in the loops' feet) is mostly likely due to collisional plasma heating by energetic precipitating particles. The direct analogy of such collisional heating and corresponding emissions are shown by optical observations of hot ($<1$ keV) plasma emissions from the top ionosphere (see Fig. \ref{fig6s}).

The interesting feature of the comparison of X-ray and EUV/UV emissions (shown in Fig. \ref{fig2f}) is that X-ray sources are much more localized, and this localization becomes even more evident for higher energies of emitting electrons. Such localization can be due to quick loss of energies of particles penetrating to the densest plasma regions, i.e. EUV (at 131 \AA{}) and soft X-ray emission ($\sim 1-10$ keV) from coronal parts of the flare loops is much less localized, than hard X-ray emission ($\sim 100$ keV) that mostly comes from the loop feet to where $\sim 100$ keV electrons may penetrate without significant losses on the way. Optical observations of emissions from the nonthermal electron ($>1$ keV) precipitations to the ionosphere also show stronger localization of such emission in comparison with emissions from $<1$ keV electron precipitations (see Fig. \ref{fig6s}). However, in absence of losses within the magnetotail, a difference between $<1$ keV and $>1$ keV emissions is mostly likely due to difference of spatial scales of mechanisms responsible for general heating ($<1$ keV) and acceleration of nonthermal ($>1$ keV) populations. Although $<1$ keV and $>1$ keV electrons are produced due to adiabatic processes of magnetic field line shrinking in the post-reconnection magnetotail \citep{Gabrielse14, Birn15}, a seed population of $<1$ keV electrons is likely subthermal background plasma (electron temperature in the magnetotail is $\sim 0.1-0.5$ keV \citep[see][]{Artemyev11:jgr} and adiabatic heating in reconnection plasma flows can produce few keV electrons  \citep[see][]{Runov15,Runov18}), whereas a seed population of $>1$ keV electrons is likely the population of electrons accelerated within the reconnection region \citep{Egedal12, Asano10, Imada07, Imada11}.  \citet{Birn17:apj} showed that a simple magnetic trap collapse (the analog of magnetic field line shrinking in the magnetotail) cannot produce nonthermal electron populations \citep[often observed as a primary source of responsible for hard X-ray emission, see, e.g., ][]{Oka13:flare} , and additional scattering and acceleration mechanisms \citep[e.g., direct acceleration by reconnection electric fields, see][]{Gordovskyy10} are needed. Thus, there is some analogy between relation of thermal and nonthermal electron production in the substorm magnetotail and in solar flares.

\subsubsection{Dynamics of Magnetic Field and Current \label{sec:flaredynamics}}

\begin{figure}[t]
\centering
\includegraphics[width=0.8\textwidth]{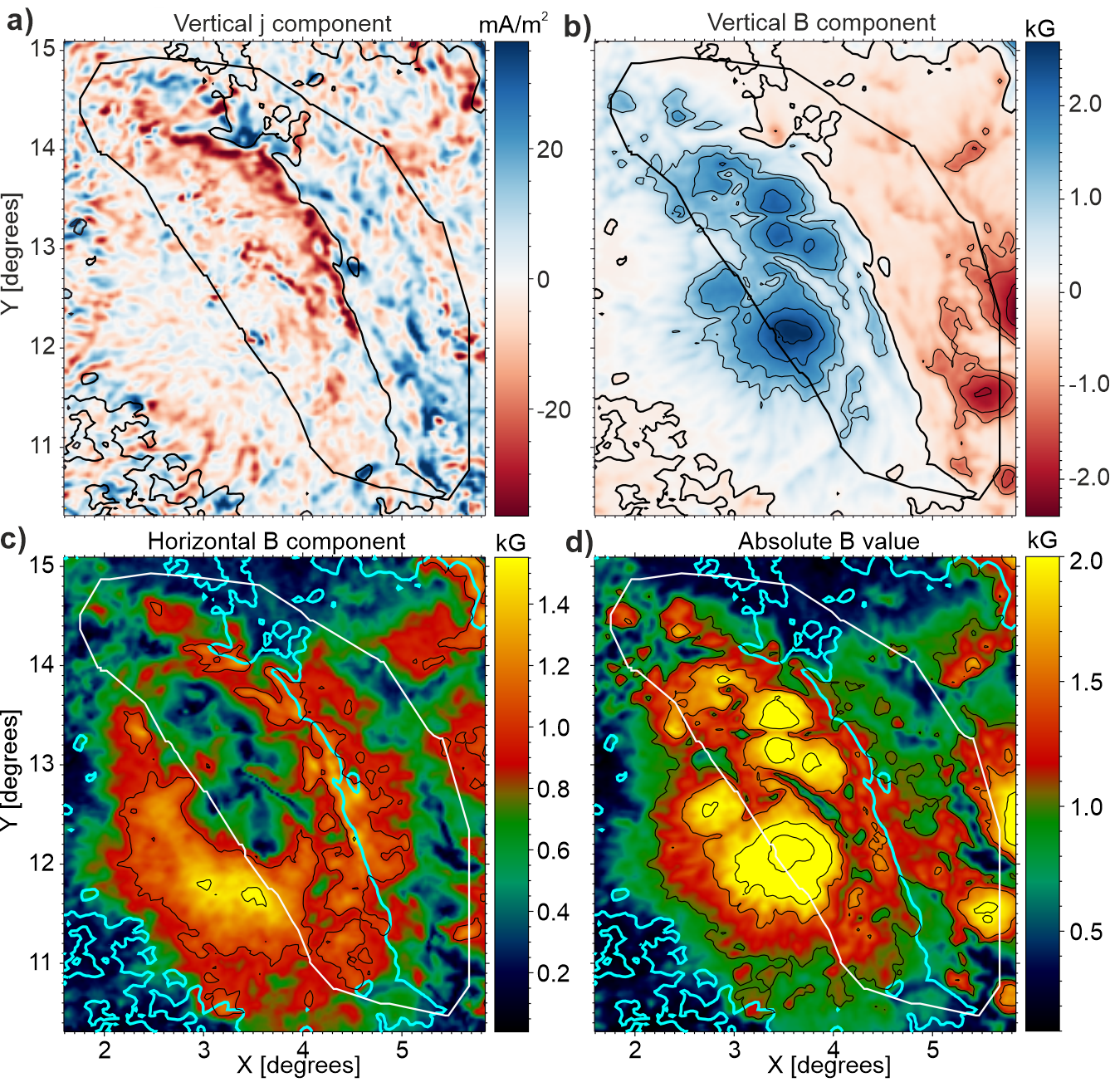}
\caption{\label{fig3f} Maps showing the distribution of vertical electric current density (a), vertical magnetic field (b), horizontal magnetic field (c), and absolute magnetic field value (d) at the photosphere deduced from the 720 s SDO/HMI vector magnetogram at around 17:36 UT, just before the {\it official} flare onset. The thin black (in (a) and (b)) and cyan (in (c) and (d)) curves show the magnetic PIL. The thick black (in (a) and (b)) and white (in (c) and (d)) closed contour indicates the outer boundary of the flare region around the PIL selected to calculate different magnetic field and electric current characteristics shown in Figure~\ref{fig4f}.}
\end{figure}

Fig. \ref{fig3f} shows spatial distributions of magnetic field components (and the absolute value of the full vector) and vertical current density at the photosphere in the active region around 3 minutes before the {\it official} flare onset (at $\approx$17:36 UT). The vertical electric current density, $j_{z}$, is calculated here from one of the Maxwell's equations (Ampere's law), using the 720 s HMI/SDO vector magnetogram. The distribution of $j_{z}$ (shown in Fig. \ref{fig3f}(a)) can be compared with the current system formed at the Earth's ionosphere and reconstructed from the ground-based magnetometer network (see Fig. \ref{fig4s}(a)). There are vertical currents of opposite polarities around the magnetic PIL, and those currents are, probably, closed, at least partially, by a return current system, not well seen in Fig. \ref{fig3f}(a) \citep[see the discussion in][]{Schmieder18GMS}. Due to presence of strong magnetic shear \citep[it can be seen as the enhanced horizontal magnetic field component around the PIL in Fig. \ref{fig2f}(c); see also][]{Wang17NatAs, Wheatland18ApJ, Kang19ApJ}, currents in the photosphere are not directly connected by cross-field currents (as opposed to the ionosphere, see Fig. \ref{fig4s}(a)), but rather connected by currents flowing almost along the PIL. Additional vertical currents formed during the solar flare (see below) should be closed at the top of the flare magnetic loops by current of energetic particles accelerated at the energy release region. Thus, there is a direct analogy between substorm and flare current systems connected cold collisional photospheric (ionospheric) plasma and hot collisionless plasma of the primary energy release cite(s).

\begin{figure}
\centering
\includegraphics[width=0.7\textwidth]{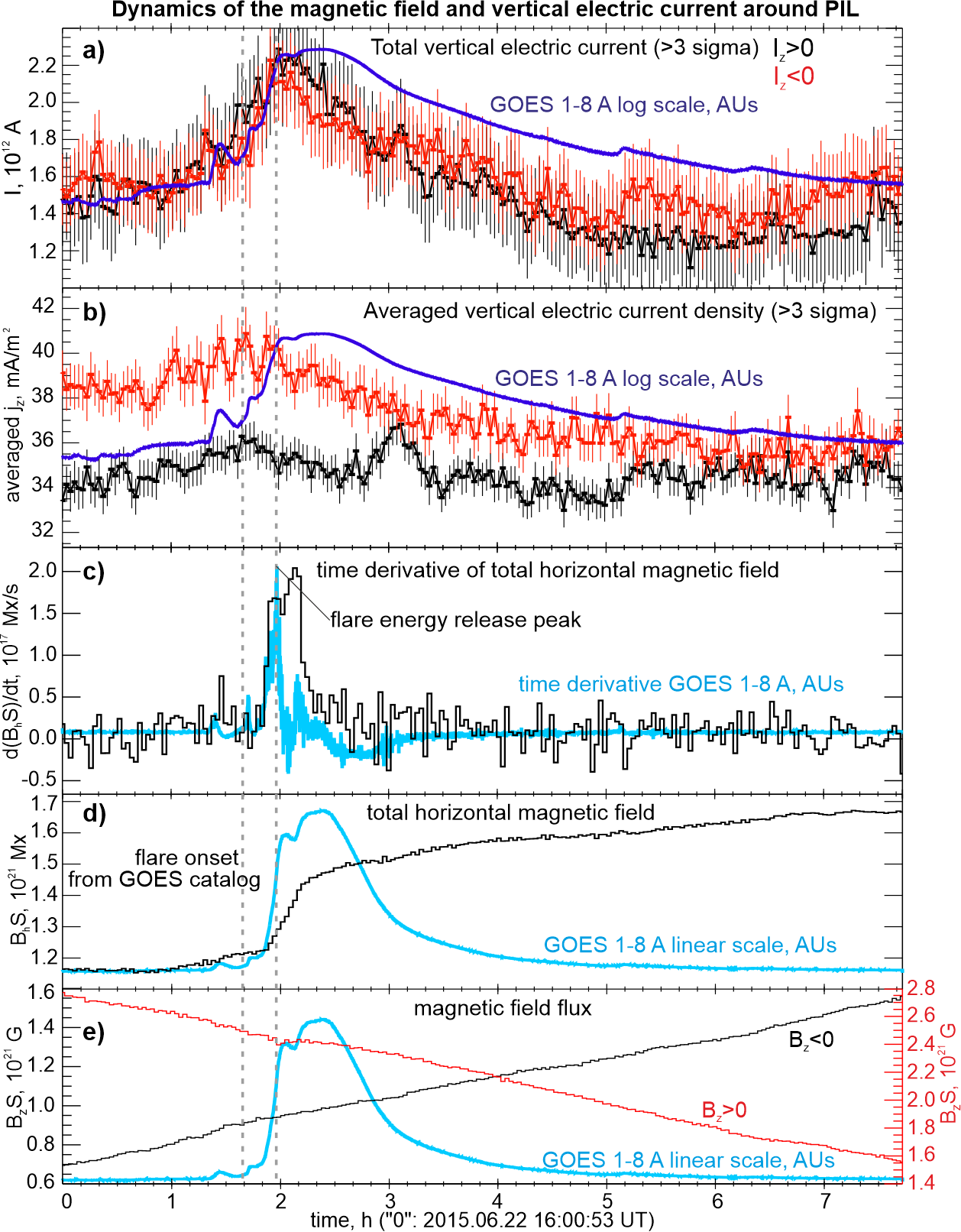}
\caption{\label{fig4f} Temporal evolution of different characteristics of magnetic field and vertical electric current at the photosphere in the flare region (marked by the thick black or white contour in Figure~\ref{fig3f}). Absolute values of the total vertical current and current density are shown in (a) and (b), respectively (positive -- black, negative -- red). Estimated errors are shown with the vertical bars. Light curve of the solar soft X-ray emission detected by GOES/XRS in the 1-8 \AA{} range is shown in the logarithmic scale (in arbitrary units) by dark blue. Time derivatives of the horizontal magnetic {\it flux} (black) and GOES 1-8 \AA{} light curve (blue) are shown in (c). Absolute values of the total horizontal and vertical magnetic {\it fluxes} are shown by black in (d) and black (for negative polarity) and red (for positive polarity) in (e), respectively. Light curve of the GOES 1–8 \AA{} in the linear scale (in arbitrary units) is shown by blue in (d) and (e). The left and right vertical dashed lines show the {\it official} flare start at 17:39 UT and the maximum of the GOES 1-8 \AA{} time derivative, respectively.}
\end{figure}

Figure \ref{fig4f}(a,b) compares dynamics of the total absolute vertical current and current density (separately for the positive and negative directions) in the flare region (shown in Fig. \ref{fig3f}) and soft X-ray emission measured by GOES/XRS. Here the vertical current is calculated from Ampere's law using the sequence of 135-s HMI/SDO vector magnetograms. A method for calculating the vertical current and estimating measurement errors is presented in \citet{Sharykin20}. We considered only current values exceeding three standard deviations ($>3 \sigma$) calculated for the background region of the Sun. There is a clear correlation between the dynamics of the total vertical current (both positive and negative) and soft X-rays emitted by plasma energized in the flare region. Thanks to the high time resolution of 135 s, one can even see the correspondence between individual spikes of X-ray emission and vertical current, in particular, during the precursors ($\sim 17:20$ UT and $\sim 17:40$ UT), as well as on the flare tail at $\sim 21:10$ UT and $\sim 22:30$ UT. This figure is an analog to Fig. \ref{fig5s} where the ionospheric currents for the substorm current system are shown to correlate well with energetic particle fluxes.

Well seen correlations of current intensities and energetic particle/soft X-ray fluxes for the substorm and flare, respectively, may suggest that there is a certain similarity between these two systems. The inflow/outflow currents on the photosphere and ionosphere should be closed to currents of energetic particles around the energy release cite(s) that is at the top of the loop or at the near-Earth magnetotail. This region is filled by energetic particles drifting across magnetic field lines (or streaming along these lines in case of the strong magnetic field shear in the solar flare) and carrying intense currents. In the magnetotail filled by hot plasma (with the plasma $\beta$ well exceeding $100$) these currents are transverse to the magnetic field, whereas in the solar flares these currents would be field-aligned or transverse (depending of the plasma $\beta$, see discussion in \cite{Gary01:solar}). The growth of these currents (and connected to them field-aligned currents measured at the photosphere/ionosphere) should be associated to the increase of energetic particle fluxes due to injections from the reconnection region (i.e., from the primary energy release cite). Observations of an increase in vertical currents in the photosphere during several flares and a discussion of their connection with reconnection in the corona can be found in, e.g., \citet{Janvier14ApJ, Janvier16A&A, Musset15A&A, Schmieder18GMS, Sharykin20}. This scenario is well consistent with in-situ spacecraft observations in the magnetotail (see Figs. \ref{fig2s}-\ref{fig4s} and \citet{Sergeev08,  Angelopoulos08,Angelopoulos13}). Thus, currents' dynamics reflect well the general plasma (heating) energization during the onset of energy release. However, the following slow decay of the current density and energetic particle fluxes may be quite different for substorms and flares. During quite long ($\sim 40$ min) period of currents/fluxes dynamics in the magnetotail, there are only two energy release events associated with strong electron precipitations (see Fig. \ref{fig5s}), whereas most of the time strong fluxes of energetic particles are trapped within the magnetotail current sheet, and drifts of these particles generate currents connected to the ionosphere through the field-aligned inflow/outflow currents shown in Fig. \ref{fig4s}. The loss-cone size in the solar flare magnetic loop is much larger than in the magnetotail, and most of energetic particles are expected to precipitate quickly after energy release. Thus, there are two possible scenarios that can explain long-term evolution (decay) of currents and soft X-ray emission in Fig. \ref{fig4f}. First, there can be continuous series of energy releases that accelerate new particles and guarantee that level of energetic particles would be sufficiently high during a long decay phase \citep[e.g.][]{Dere79ApJ, Zimovets12SoPh, Qiu16ApJ, Yu20ApJ}. Second, there can be strong magnetic turbulence trapped initially energized particles around the top of the flare loops \citep[see discussion in][]{Karlicky&Kosugi04,Grady&Neukirch09}. The second scenario is much closer to the substorm system with few energy releases and long life-time of energetic fluxes and currents.

Figure \ref{fig4f}(d) compares dynamics of the horizontal magnetic {\it flux} (i.e. the horizontal magnetic field component multiplied by pixel area and integrated over the flare region containing the inflow/outflow or negative/positive currents shown in Fig. \ref{fig3f}(a)) and soft X-ray emission flux in the GOES/XRS 1-8 \AA{} channel. The increase of the horizontal magnetic flux correlates well with soft X-ray flux produced by hot plasma in the flare region. This correlation is also clearly visible in Fig. \ref{fig4f}(c), which shows the time derivatives of the horizontal magnetic flux and soft X-ray flux. At the same time, it can be noted that the fluxes of the vertical magnetic field changes relatively smoothly before, during and after the flare, and does not show a jump during the growth of X-ray flux (Fig. \ref{fig4f}(e)). Step-like increase in the horizontal component of the magnetic field in the flare region near the PIL is also detected in other flares \citep[e.g.][]{Petrie13SoPh, Sun17ApJ, Sharykin20}. There is a direct analogy with the magnetic flux increase in the near-Earth region shown in Fig. \ref{fig5s}. For the Earth's magnetotail such magnetic field increase is due to magnetic flux transport from the energy release region to the near-Earth side, where strong dipole field breaks the plasma flow \citep[for details of such flux increase correlation with the energetic electron precipitations see, e.g.,][]{Gabrielse19}. Around the flow breaking region the transported magnetic flux is accumulated and there so-called magnetic-pilled-up region is formed. This region expands toward the magnetotail and such and expansion can switch-off the magnetic reconnection \citep{Baumjohann02}. Strong magnetic field within this region traps energetic particles \citep[e.g.,][and references therein]{Turner16, Gabrielse17}. Oscillations of the outer boundary (edge) of this region is seen on MMS spacecraft as quasi-periodic increase of the equatorial, $B_z$ (it corresponds to $B_{h}$ for the solar flare), magnetic field (see Fig. \ref{fig2s}). Does the same mechanism of energetic particle trapping work for solar flares and can this mechanism substitute/supplement trapping by magnetic field turbulence? Although the horizontal magnetic flux increase shown in Fig. \ref{fig4f}(c,d) indicates such similarity, additional confirmations are needed. A possible dataset of observations that would show energetic particle trapping due to magnetic field increase at the top of the loops is the spatially localized and spectrally resolved gyrosynchrotron emission that is presumably generated within the regions of enhanced magnetic field. In some flares, there is indeed increased microwave emission at the tops (or above them) of the flare loops, indicating efficient electron trapping in these regions \citep[e.g.][]{Melnikov02ApJ, Minoshima08ApJ, Huang09ApJ, Reznikova09ApJ}. However, these studies did not establish temporal changes in the magnetic field at the loop-tops during the flares studied. The accumulation of energetic electrons at the loop-tops was interpreted as trapping due to the specific anisotropic pitch-angle distributions of the injected accelerated electrons. In some works, the possible influence of turbulence on the efficiency of trapping of energetic electrons in flare loops was discussed \citep[e.g.][]{Charikov&Shabalin15, Filatov20Ge&Ae}. Most recently, \citet{Fleishman20}, based on the analysis of spatially and spectrally resolved microwave observations with EOWSA, showed a simultaneous decrease in the magnetic field (and magnetic energy density) with an increase in the energy density of nonthermal electrons in the region above the flare loops in the solar flare on September 10, 2017. This result was interpreted by transformation of magnetic energy into kinetic energy of accelerated electrons due to magnetic reconnection within the {\it standard} eruptive flare model. However, we are not aware of studies in which a simultaneous increase in the longitudinal ($B_{h}$) component of the magnetic field at the top of the flare loops and an increase in the density of nonthermal electrons would be shown in the same place.

\section{Discussion and Conclusions \label{sec:discussion}}
This section aims to discuss comparisons of datasets presented for the magnetotail substorm and solar flare. We do not intend to compare these datasets directly, but rather we would like to consider substorm dataset to suggest some possible scenario for interpretation of datasets for solar flare. We discuss three main subtopics of this study, show table demonstrating rough analogies between different datasets, and finally conclude our investigation.

\subsection{Formation of substorm/flare current system}
Both magnetotail substorm and solar flare demonstrate formation of the system of intense field-aligned currents that connect the region of hot plasma (energy release region) and region of dense (conductive) cold plasma (ionosphere or photosphere). In the Earth magnetotail such system originates from the dynamics of plasma flows generated at the energy release region and breaking in the near-Earth strong dipole field region \citep[see simulations by][]{Birn11,Birn&Hesse13:scw,El_Alaoui13}. The main equation describing field-aligned current closer to the cross-field currents within the flow breaking region is the divergence free equation of currents \citep{Vasyliunas70}, i.e. the basic physics of such field-aligned current system is well described within MHD models and should be well reproduced in global MHD simulations of solar flares. Assuming this similarity between these systems in substorm magnetotail and solar flares, we can make several comments about kinetics of field-aligned currents. From the magnetotail observations we know that, being generated as a global MHD system of field-aligned currents, this system quickly involves ion-scale and electron-scale kinetics \citep[see reviews by][]{Lysak90, Stasiewicz00}. The most relevant to the solar flares, the filamentation of field-aligned currents due to transformation of Alfven waves (carrying such currents) to kinetic or inertial Alfven waves \citep{Lysak04,Lysak13,Chaston14}. Such transformation is quite important, because it opens a door for collisionless dissipation of currents by thermal electrons accelerated by field-aligned electric fields of kinetic/inertial Alfven waves \citep[see discussion of these processes in application to the solar corona in, e.g., ][]{Fletcher&Hudson08, Haerendel12,Artemyev16A&A}. Therefore, some elements of kinetic physics from substorm observations (e.g., efficiency of electron acceleration/field-aligned current damping, see \citet{Damiano16,Pyakurel18:kaw}) may be implemented into solar flare models. Moreover, ground-based observations of substorms  show a direct relation between enhanced horizontal currents and electric fields around the field-aligned currents \citep{Opgenoorth83}. Such observations may be useful for investigation and simulations of electric fields on the photosphere level \citep[see discussion of this analogy in][]{Kropotkin11:asp, Krasnoselskikh12:conf}. 

For the interpretation of changes in the magnetic fields and vertical currents in the investigated flare of June 22, 2015, \citet{Wheatland18ApJ} have proposed a simplified analytical model based on the response of the photosphere to a large-amplitude Alfven wave propagating downward from the energy release cite in the corona. The wave brings magnetic and velocity shear into the photosphere, which can qualitatively explain the observed changes. \citet{Wheatland18ApJ} also raised the issue that electrons can be accelerated in the field-aligned electric field at the front of this large-scale Alfven wave. To fulfill the condition that this field-aligned electric field exceeds the Dreicer field (when electrons can be effectively accelerated in the runaway mode), the front width should be rather narrow, $\approx 10$ m. However, the authors mentioned that in the presence of an anomalous resistivity due to turbulence or micro-scale structures, the required front thickness could be much larger. Although large-scale Alfven wave transformation into kinetic/inertial Alfven waves was not considered by \citet{Wheatland18ApJ}, the general issue of energy cascade from the large to small scales in Alfven wave turbulence is quite similar for both magnetotail and solar corona systems.

\subsection{Energetic particle fluxes and their correlation with currents}
Both substorm magnetotail and solar flare demonstrate a correlation of energetic particle fluxes/X-ray emission with the dynamics of electric currents. This correlation is well explained and modeled for the magnetotail, where ionospheric currents are closed through the magnetotail currents generated by energetic particle drifts \citep{Kepko14:ssr, Ganushkina15}. However, this explanation is based on the concept that energetic particles are trapped in the near-Earth region for a long time, and their precipitations into the narrow loss-cone (few degrees) are driven by weak pitch-angle diffusion due to curvature scattering (working mostly for energetic ions, see \citet{Sergeev12:IB, Sergeev15}, but contributing to electron losses as well, see \cite{Eshetu18,Artemyev13:angeo:scattering}) and wave turbulence (working mostly for energetic electrons, see \citet{Nishimura10:Science, Nishimura20:ssr, Ni16:ssr}). Also field-aligned currents can be driven by pressure gradient and flow shear. In the solar flare loops with the large loss-cone \cite{Eradat14} such long trapping of energetic particles would require pitch-angle scattering away from the loss-cone, i.e., would require strong turbulence level (see discussion in \citet{Hannah&Kontar11, Fleishman18, Fleishman20}). Alternatively, energetic particle population can be continuously supplemented by series of transient energy releases. The first scenario would resemble a weak isolated magnetotail substorm (shown in this study), whereas the second scenario is an analog of multiple substorms occurring within strong geomagnetic storm \cite[see, e.g.,][]{Gabrielse17,Angelopoulos20}. 

If we assume that the correlation of energetic fluxes and currents has similar origins in substorms and flares, we can speculate on importance of elements of substorm physics for solar flares. First, strong drifts (different for ion and electrons) of energetic particles trapped in the near-Earth magnetotail  result in electric polarization of this region. Such electric fields are projected along magnetic field lines to the ionosphere and are observed as ionospheric plasma drifts \citep{Sergeev04:angeo}. Similar observations of photosphere cross-field plasma motions \citep[see discussion in, e.g.,][]{Krasnoselskikh10:apj} may be interpreted in terms of electric fields dynamics in analogy to the ionosphere-magnetosphere coupling investigations. Second, energetic particle populations in the near-Earth magnetotail create a strong pressure gradient unstable to various drifts modes. Plasma and field perturbations by unstable waves are well seen in in-situ measurements \citep[e.g.,][]{Panov12:grl, Panov13}, numerical simulations \citep[e.g.,][]{Pritchett10, Pritchett14, Sorathia20}, and ground-based optical observations \citep[e.g.,][]{Nishimura16}. Similar waves are expected to be developed on the top of flare loops, around the breaking of plasma flows from the energy release region. Properties of such waves, generated by pressure gradients of energetic particles, may be used for estimates of characteristics energetic particle populations. Indeed, oscillations and/or quasi-periodic pulsations (QPPs) are observed at least in some flares in the vicinity of the loop-tops, where nonthermal electrons are present \citep[e.g.][]{Jakimiec10SoPh,Yuan19ApJ,Reeves20ApJ}. The nature of these oscillations is not yet clear. According to the MHD model developed by \citet{Takasao16ApJ}, these oscillations can result from the formation a {\it magnetic tunning fork} structure at the tops of reconnected loops due to their interaction with accelerated plasma flows falling from the reconnection region from above. However, other mechanisms of these oscillations cannot be ruled out so far, and research should be continued, in particular, taking into account the aforementioned physical processes established for magnetospheric substorms.

\subsection{Thermal versus nonthermal electron populations}
Both substorm magnetotail and solar flare demonstrate difference in spatial localizations of emissions associated with precipitations of energetic (several tens of keV and more) and hot ($\leq 1$ keV) electrons. Optical observations from the ground-based all sky cameras show that emission due to $>1$ keV electron precipitations to the ionosphere is much more spatially localized and temporally transient than emissions due to $<1$ keV precipitations. Similarly, hard X-ray and microwave emissions are more bursty than soft X-ray and UV/EUV emissions (see Fig. \ref{fig1f}), and the hard X-ray sources are usually much more localized than the extended flare ribbons observed in UV and optical (e.g. H$_{\alpha}$) ranges and soft X-ray and EUV flare loops rooted in these ribbons \citep[see Fig. \ref{fig2f}, and also, e.g., ][]{Temmer07ApJ,Fletcher11}. For the substorm magnetotail such localization can be explained by difference of mechanisms responsible for electron heating (below $\sim 1$ keV) and acceleration (to tens of keV). Simulations \citep[e.g.,][]{Birn15,Gabrielse16,Gabrielse17} and observations \citep[e.g.,][]{Runov15} show that the main energization occurs within fast plasma flows (especially at the flow front, see \citet{Sergeev09,Runov11jgr}) originated from the energy release region. These flows are responsible for magnetic field compression and field line shrinking that adiabatically energize the main electron populations, i.e. energization process being adiabatic depends on the initial electron energy. Energization of the background electron population should result in heating from $0.1$ keV to few keVs for flows traveling from the middle tail reconnection to the near-Earth flow breaking region \citep[e.g.,][]{Runov14,Gabrielse17}. However, the energy release (and associated electric field bursts, see \citet{Egedal12, Torbert18,Angelopoulos20}) also produce accelerated electron population that can be further heated to tens of keV in the magnetotail \citep{Runov13,Fu13:NatPh}. This population, being produced within a specially localized region of the transient magnetic reconnection \citep{Turner21:grl, Cohen21:grl}, would be much more localized than the heated background electron populations. Similar logic may be adopted for solar flares. Heating of the main electron population by energetic electron precipitations generates suprathermal electrons (an analog of hot background magnetotail population) that may be adiabatically heated to energies sufficient for soft X-ray emissions, and these heated particles may heat the plasma in the transition region and chromosphere responsible for the flare ribbons, whereas only electrons pre-accelerated within the energy release region can be further adiabatically heated to $\gtrsim 30$ keV energies associated with the footpoint hard X-ray sources. Population of such energetic electrons is much smaller \citep[by their number, but not by their total energy; see e.g.][]{Emslie12ApJ,Aschwanden16ApJ}, and their emission is seen only from the densest plasma region around the loop feet. Such an substorm/flare analogy indicates on possible importance of adiabatic processes for electron heating in solar flares \citep[see also][]{Bogachev07, Borissov16, Birn17:apj}, because these processes are mainly responsible for energization of the magnetotail plasma.

\begin{table}[b]
  \begin{tabular}{ | p{4cm} | p{4cm} | p{9cm} | }
    \hline
    solar flare & magnetosphere substorm & comments \\ \hline\hline
    photosphere magnetic field and reconstructed vertical currents & ground-based field and reconstructed currents & similar dynamics along the substorm/flare and similar closure to the energetic particle currents around energy release region  \\ \hline
    hard X-ray emission from the loop roots & riometer signals & direct indication of $>30$ keV electron precipitations to the chromosphere/ionosphere from the energy release region \\\hline
    hard X-ray emissions from the heated loops & in-situ measurements of energetic fluxes & similar correlation to current enhancements and direct relation to plasma energization by energy release \\\hline
    soft X-ray and EUV emissions from the heated loops & optical measurements of energetic electron precipitations & similar emission generation mechanisms from hot and energetic electron precipitating to the chromosphere/ionosphere \\\hline   
  \end{tabular}
  \caption{Dataset comparison for solar flares and magnetosphere substorms}\label{table1}
\end{table}

\subsection{Conclusions}
In this study we compare several datasets available for the solar flares and for magnetosphere substorms. Similarities of observations in terms of measured quantities or energy range of particles, illustrated on the comparative analysis of one typical substorm and one eruptive flare, are summarized in Table \ref{table1}. The main similarity discussed in this study is the common relation of current and energetic fluxes dynamics: rapid increase of inflow/outflow currents on the photosphere and ionosphere well correlates with energetic electron generation and precipitations (as shown by hard X-ray emissions and riometer signals), whereas long-term decay of currents correlates with dynamics of energetic electron fluxes filled flare loops and near-Earth magnetotail (as shown by soft X-ray emissions and in-situ plasma measurements in the magnetotail). Optical observations from the all-sky camera and X-ray (together with UV/EUV) emission for separated energies demonstrate the similar effect of more spatially localized precipitations from higher energy electrons.


\section*{Acknowledgements}
Used MMS, THEMIS, and ERG data can be downloaded from http://lasp.
colorado.edu/mms/sdc, http://themis.ssl.berkeley.edu, respectively. Science data of the ERG (Arase) satellite were obtained from the ERG Science Center operated by ISAS/JAXA and ISEE/Nagoya University (https://ergsc.isee.nagoya-u.ac.jp/index.shtml.en, \citep{Miyoshi18:ERGdata}). 

NRCan riometer data used in this study may be obtained in \url{https://doi.org/10.7910/DVN/CHD02S} or by the request from Robyn Fiori (Robyn.Fiori@canada.ca). Operational support for NORSTAR is provided by the Canadian Space Agency through the GO Canada program. NORSTAR data were made available directly from the University of Calgary and may be freely obtained from the Canadian Space Science Data Portal (http://www.cssdp.ca).

Data analysis was done using SPEDAS V3.1, see \citet{Angelopoulos19}. We are grateful to the teams of SDO/HMI, SDO/AIA, RHESSI, GOES/XRS, Fermi/GBM, and RSTN for the observational data provided. 

The work is supported by the Russian Science Foundation grant No. 17-72-20134 (IZ, IS; Sec. \ref{sec:flare}), NASA contract NAS5‐02099 (AA, AR, VA, XZ), NASA DRIVE HERMES grant No. 80NSSC20K0604 (AA, AR, YN, CD, MV, OP), NASA grant 80NSSC18K0657, 80NSSC20K0725, NSF grant AGS-1907698, and AFOSR grant FA9559-16-1-0364 (YN).

The all-sky imager at South Pole is supported by NSF ANT-1643700 and Japanese Antarctic Research Expedition program.



\end{document}